\documentclass{w-book}

\newcommand{\minrm}{{\rm min}}
\newcommand{\maxrm}{{\rm max}}
\newcommand{\frm}{{\rm f}}

\newcommand{\exprm}{{\rm exp}}

\newcommand{\drm}{{\rm d}}
\newcommand{\grm}{{\rm g}}

\newcommand{\X}{{\rm X}}

\newcommand{\pa}{\partial}

\newcommand{\text}{\rm}

\newcommand{\ug}{ \; = \; }
\newcommand{\ugg}{ \ = \ }

\newcommand{\infi}{\infty}

\newcommand{\kr}{k_{\rho}}

\newcommand{\al}{\alpha}
\newcommand{\ze}{\zeta}

\newcommand{\bb}{\begin{equation}}
\newcommand{\ee}{\end{equation}}
\newcommand{\bega}{\begin{eqnarray}}
\newcommand{\ega}{\end{eqnarray}}
\newcommand{\begae}{\begin{eqnarray*}}
\newcommand{\egae}{\end{eqnarray*}}

\newcommand{\h}{\hspace*{4ex}}
\newcommand{\dis}{\displaystyle}

\newcommand{\be}{\beta}

\newcommand{\om}{\omega}

\newcommand{\wmin}{\om_{{\rm min}}}
\newcommand{\wmax}{\om_{{\rm max}}}

\usepackage[dvipsone]{graphicx}
\usepackage{w-bookps}

\usepackage{cite}

\begin{document}


\booktitle{} \subtitle{}

\author{}

\baselineskip 0.5cm  

\vspace{1cm}

{\Large{ Structure of the Nondiffracting (Localized) Waves, and
some interesting applications} }

\vspace{1cm}

Michel Zamboni-Rached\\
\textit{\small Centro de Ci\^encias Naturais e Humanas,} \\
\textit{\small Universidade Federal do ABC, Santo Andre, SP, Brasil}\\
\\
Erasmo Recami\\
\textit{\small Facolt\`a di Ingegneria, Universit\`a statale di
Bergamo, Bergamo, Italy; and} \\
\textit{\small INFN---Sezione di Milano, Milan, Italy}\\
\\
Hugo E. Hern\'{a}ndez-Figueroa\\
\textit{\small DMO--FEEC, State University at Campinas, Campinas,
SP, Brazil} \\



\section[Introduction]{Introduction}\label{sec:Introduction}

Since the early works[1-4] on the so-called nondiffracting waves
(called also Localized Waves), a great deal of results has been
published on this important subject, from both the theoretical and
the experimental point of view. Initially, the theory was developed
taking into account only free space; however, in recent years, it
has been extended for more complex media exhibiting effects such as
dispersion[5-7], nonlinearity[8], anisotropy[9] and losses[10].
Such extensions have been carried out along with the development of
efficient methods for obtaining nondiffracting beams and pulses in
the subluminal, luminal and superluminal regimes[11-18].
This paper (mainly a review) addresses some theoretical methods related
to nondiffracting solutions of the linear wave equation in unbounded
homogeneous media, as well as to some interesting applications of
such waves.

The usual cylindrical coordinates $(\rho,\phi,z)$ will be used
here. In these coordinates the linear wave equation is written as

\bb
\frac{1}{\rho}\frac{\pa}{\pa\rho}\left(\rho\frac{\pa\Psi}{\pa\rho}
\right) + \frac{1}{\rho^2}\frac{\pa^2\Psi}{\pa\phi^2} +
\frac{\pa^2\Psi}{\pa z^2} - \frac{1}{c^2}\frac{\pa^2\Psi}{\pa t^2}
\ug 0 \label{eo}\ee

In section II we analyze the general structure of the Localized
Waves, develop the so called Generalized Bidirectional
Decomposition, and use it to obtain several luminal and
superluminal (especially X-shaped) nondiffracting wave solutions
of eq.(\ref{eo}).

In section III we develop a space-time focusing method
by a continuous superposition of X-Shaped pulses of different
velocities.

Section IV addresses the properties of chirped optical X-Shaped
pulses propagating in material media without boundaries.

Finally, in Section V, we show how a suitable superposition of
Bessel beams can be used to obtain stationary localized wave
fields with a static envelope and a high transverse
localization, and whose longitudinal intensity pattern can
assume any desired shape within a chosen
interval $0 \leq z \leq L$ of the propagation axis.

\section{Spectral structure of the Localized Waves and the Generalized
Bidirectional Decomposition}\label{secII}

An effective way to understand the concept of the (ideal)
nondiffracting waves is furnishing a precise mathematical definition
of these solutions, so to extract the necessary spectral
structure from them.

Intuitively, an ideal nondiffracting wave (beam or pulse) can be
defined as a wave capable of maintaining indefinitely its spatial
form (apart from local variations) while propagating.

We can express this intuitive property saying that a localized
wave has to possess the property[12,13]

\bb \Psi(\rho,\phi,z,t) \ug \Psi(\rho,\phi,z + \Delta z_0,t +
\frac{\Delta z_0}{V}) \label{S2def} \ee where $\Delta z_0$ is a
certain length and $V$ is the pulse propagation speed that here
can assume any value: $0\leq V \leq \infty$.

Using a Fourier Bessel expansion, we can express a function
$\Psi(\rho,\phi,z,t)$ as

\bb \Psi(\rho,\phi,z,t) \ug
\sum_{n=-\infty}^{\infty}\left[\int_{0}^{\infty}d\kr\,\int_{-\infi}^{\infty}dk_z\,\int_{-\infi}^{\infty}d\om\,\kr
A_n^{'}(\kr,k_z,\om) J_n(\kr \rho)e^{ik_z z}e^{-i\om t}e^{i n
\phi} \right] \, . \label{S2geral1} \ee

Using the translation property of the Fourier transforms
$\emph{T}[f(x+a)] = {\rm exp}(ika)\emph{T}[f(x)]$, we have that
$A_n^{'}(\kr,k_z,\om)$ and ${\rm exp}[i(k_z\Delta z_0 -\om\Delta
z_0/V)]A_n^{'}(\kr,k_z,\om)$ are the Fourier Bessel transforms of
the l.h.s and r.h.s. functions in eq.(\ref{S2def}). And from this
same equation we can get[12,13] the fundamental constraint linking
the angular frequency $\om$ and the longitudinal wavenumber $k_z$:

\bb \om \ug Vk_z + 2m\pi \frac{V}{\Delta z_0} \label{S2cond2} \ee
with $m$ an integer. Obviously, this constraint can be satisfied
through the spectral functions $A_n^{'}(\kr,k_z,\om)$.

Now, let us explicitly mention that constraint (\ref{S2cond2})
does not imply any breakdown of the wave equation validity. In
fact, when inserting expression (\ref{S2geral1}) in the wave
equation (\ref{S2eo}), one gets that

\bb \frac{\om^2}{c^2} \ug k_z^2 + \kr^2 \label{S2cond3}  \ee

So, to obtain a solution of the wave equation from
(\ref{S2geral1}), the spectrum $A_n^{'}(\kr,k_z,\om)$ must have the
form

\bb A_n^{'}(\kr,k_z,\om) \ug A_n(k_z,\om)\,\delta\left(\kr^2 -
\left(\frac{\om^2}{c^2} - k_z^2 \right)\right) \label{S2sepc1} \ee
where $\delta(.)$ is the Dirac delta function. With this we can
write a solution of the wave equation as

\bb \Psi(\rho,\phi,z,t) \ug
\sum_{n=-\infty}^{\infty}\left[\int_{0}^{\infty}d\om\,\int_{-\om/c}^{\om/c}dk_z
A_n(k_z,\om) J_n\left(\rho\sqrt{\frac{\om^2}{c^2} - k_z^2}
\right)e^{ik_z z}e^{-i\om t}e^{i n \phi} \right] \label{S2geral2}
\ee where we have considered positive angular frequencies only.

Equation (\ref{S2geral2}) is a superposition of Bessel beams and
it is understood that the integrations in the $(\om,k_z)$ plane
are confined to the region $0\leq \om \leq \infty$ and $-\om/c
\leq k_z \leq \om/c$

Now, to obtain an ideal nondiffracting wave, the spectra
$A_n(k_z,\om)$ must obey the fundamental constraint
(\ref{S2cond2}),  and so we write

\bb A_n(k_z,\om) \ug
\sum_{m=-\infty}^{\infty}\,S_{nm}(\om)\delta\left(\om - (Vk_z +
b_m) \right) \label{S2speclw} \ee

where $b_m$ are constants representing the terms $2m\pi V/\Delta
z_0$ in eq.(\ref{S2cond2}), and $S_{nm}(\om)$ are arbitrary
frequency spectra.

With eq.(\ref{S2speclw}) into eq.(\ref{S2geral2}), we get a general
integral form of an ideal nondiffracting wave defined by
eq.(\ref{S2def}):

\bb \Psi(\rho,\phi,z,t) \ug
\sum_{n=-\infty}^{\infty}\,\sum_{m=-\infty}^{\infty}
\psi_{nm}(\rho,\phi,z,t) \label{S2suppsinm}\ee

with

\bb \begin{array}{clcr} \psi_{nm}(\rho,\phi,z,t) \ug &
\dis{e^{-ib_m
z/V}\,\int_{(\wmin)_m}^{(\wmax)_m}\, d\om \, S_{nm}(\om)} \\

\\

& \times \dis{
J_n\left(\rho\sqrt{\left(\frac{1}{c^2}-\frac{1}{V^2}\right)\om^2 +
\frac{2b}{V^2}\om - \frac{b^2}{V^2}}
\right)e^{i\frac{\om}{V}(z-Vt)}e^{i n \phi}} \end{array}
\label{S2psinm} \ee where $\wmin$ and $\wmax$ depend on the values
of V:

\begin{itemize}
    \item{For subluminal $(V<c)$ localized waves: $b_m>0$,
    $(\wmin)_m=cb_m/(c+V)$ and $(\wmax)_m=cb_m/(c-V)$}.
    \item{For luminal $(V=c)$ localized waves: $b_m>0$,
    $(\wmin)_m=b_m/2$ and $(\wmax)_m=\infty$}.
    \item{For superluminal $(V>c)$ localized waves: $b_m\geq 0$,
    $(\wmin)_m=cb_m/(c+V)$ and $(\wmax)_m=\infty$. Or $b_m<0$,
    $(\wmin)_m=cb_m/(c-V)$ and $(\wmax)_m=\infty$}.
\end{itemize}

It is important to notice that each $\psi_{nm}(\rho,\phi,z,t)$ in
the superposition (\ref{S2suppsinm}) is a truly nondiffracting
wave (beam or pulse) and the superposition of them,
(\ref{S2suppsinm}), is just the most general form to represent a
nondiffracting wave defined by eq.(\ref{S2def}). Due to this fact,
the search for methods capable of providing analytical solutions for
$\psi_{nm}(\rho,\phi,z,t)$, eq.(\ref{S2psinm}), becomes an
important task.

Let us remember that equation (\ref{S2psinm}) is also a Bessel
beam superposition, but with the constraint (\ref{S2cond2})
between their angular frequencies and longitudinal wavenumbers.

In spite of the fact that the expression (\ref{S2psinm})
represents ideal nondiffracting waves, it is difficult to obtain
closed analytical solutions from it. Due to this, we are going to
develop a method capable of overcoming this limitation, providing
several interesting localized wave solutions (luminal and
superluminal) of arbitrary frequencies, including some solutions
endowed with finite energy.

\subsection{The Generalized Bidirectional Decomposition}

For reasons that will be clear soon, instead of dealing with the
integral expression (\ref{S2suppsinm}), our starting point is the
general expression (\ref{S2geral2}).

Here, for simplicity we will restrict ourselves to axially
symmetric solutions, assuming the spectral functions so
that

\bb A_n(k_z,\om) \ug \delta_{n0}A(k_z,\om) \ee where $\delta_{n0}$
is the Kronecker delta.

In this way, we get the following general solution (considering
positive angular frequencies only) which describes axially symmetric
waves:

\bb \Psi(\rho,\phi,z,t) \ug
\int_{0}^{\infty}d\om\,\int_{-\om/c}^{\om/c}dk_z A(k_z,\om)
J_0\left(\rho\sqrt{\frac{\om^2}{c^2} - k_z^2}\right)e^{ik_z
z}e^{-i\om t}\label{S2geral4} \ee

As we have seen, ideal nondiffracting waves can be obtained since
the spectrum $A(k_z,\om)$ satisfies the linear relationship
(\ref{S2cond2}). In this way, it is natural to adopt new spectral
parameters in the place of $(\om,k_z)$ that make easier to
implement that constraint[12,13].

With this in mind, we choose the new spectral parameters
$(\al,\be)$ BY

\bb \al \ug \frac{1}{2V}(\om + Vk_z) \;;\;\;\; \be \ug
\frac{1}{2V}(\om - Vk_z) \label{S2ab}\ee

\emph{Let us consider here only luminal $(V=c)$ and superluminal
$(V>c)$ nondiffracting pulses}.

With the change of variables (\ref{S2ab}) in the integral
solution(\ref{S2geral4}), and considering $(V \geq c)$, the
integration limits on $\al$ and $\be$ have to satisfy the three
inequalities

\bb \left\{ \begin{array}{clcr} 0 < \al + \be < \infty \\

\\

\al \geq \dis{\frac{c-V}{c+V}}\be   \\

\\

\al \geq \dis{\frac{c+V}{c-V}}\be \end{array}\right.\label{S2inq2}
\ee

Let us suppose both $\al$ and $\be$ to be positive $[\al,\,\be\geq
0]$. The first inequality in (\ref{S2inq2}) is then satisfied;
while the coefficients $(c-V)/(c+V)$ and $(c+V)/(c-V)$ entering
relations (\ref{S2inq2}) are both negatives (since $V > c$). As a
consequence, the other two inequalities in (\ref{S2inq2}) result
to be automatically satisfied. In other words, the integration
limits in $0\leq \al \leq \infty$ and $0\leq \be \leq \infty$ are
\emph{contained} in the limits (\ref{S2inq2}) and are therefore
acceptable. Indeed, they constitute a rather suitable choice for
facilitating all the subsequent integrations.

Therefore, instead of eq.(\ref{S2geral4}), we shall consider the
(more easily integrable) Bessel beam superposition in the new
variables [with $V \geq c$]

\bb\Psi(\rho,\ze,\eta) \ug \int_{0}^{\infi}
 d\al \int_{0}^{\infi}d\be \,A(\al,\be)\,
J_0\left(\rho\,\sqrt{\left(\frac{V^2}{c^2}-1\right)
 (\al^2+\be^2)+2\left(\frac{V^2}{c^2}+1\right)\al\be}\, \right) \,
e^{i\al\ze}\,e^{-i\be\eta}\label{S2geral5}\ee where we have
defined

\bb \zeta \equiv z-Vt\;;\;\;\;\;\;\;\; \eta \equiv z+Vt \ee

The present procedure is a generalization of the so-called
``bidirectional decomposition" technique[11], which was devised in
the past for $V = c$.

\emph{From the new spectral parameters defined in transformation
(\ref{S2ab}), it is easy to see that the constraint
(\ref{S2cond2}), i.e. $\om = Vk_z + b$, is implemented just by
making}

\bb A(k_z,\om) \rightarrow  A(\al,\be) \ug S(\al)\delta(\be-\be_0)
\label{S2specab} \ee with $\beta_0=b/2V$. The delta function
$\delta(\be-\be_0)$ in the spectrum (\ref{S2specab}) means that we
are integrating Bessel beams along the continuous line $\om = Vk_z
+ 2V\be_0$ and, in this way, the function $S(\al)$ will give the
frequency dependence of the spectrum: $S(\al)\rightarrow S(\om/V -
\be_0)$.

\emph{This method is a natural way of obtaining pulses with field
concentration on $\rho=0$ and $\zeta=0 \rightarrow z=Vt$.}

Now, it is important to stress[13] that, when $\be_0>0$ in
(\ref{S2specab}), the superposition (\ref{S2geral5}) has
contributions from both backward and forward Bessel beams in the
frequency intervals $V\be_0 \leq \om < 2V\be_0$ (where $k_z<0$)
and $2V\be_0 \leq \om \leq \infty$ (where $k_z\geq 0$),
respectively.

Nevertheless, we can obtain physical solutions when making the
contribution of the backwards components negligible, by choosing
suitable weight functions $S(\al)$.

It is also important to notice that we use the new spectral
parameters $\al$ and $\be$ just to obtain (closed-form) analytical
localized wave solutions, IT being that the spectral characteristics
of these new solutions can be brought into evidence just by using
transformations (\ref{S2ab}) and writing the corresponding
spectrum in terms of the usual $\om$ and $k_z$ spectral
parameters.

In the following, we consider some cases with $\be_0=0$ and
$\be_0>0$.

\

\subsubsection{Closed analytical expressions describing some ideal
nondiffracting pulses}

\

\

\emph{Let us first consider, in eq.(\ref{S2geral5}), the following
spectra, of the type (\ref{S2specab}) with $\be_0 = 0$:}

\bb A(\al,\be) \ug aV\,\delta(\be)e^{-aV\al} \label{S2spec1} \ee

\bb A(\al,\be) \ug aV\,\delta(\be)J_0(2d\sqrt{\al})e^{-aV\al}
\label{S2spec2} \ee

\bb A(\al,\be) \ug \delta(\be)\frac{\sin(d\al)}{\al}e^{-aV\al}
\label{S2spec4} \ee with $a>0$ and $d$ being constants.

One can obtain from the above spectra the following superluminal
localized wave solutions, respectively:

--- From spectrum (\ref{S2spec1}), we can use the identity (6.611.1)
in ref.[19], obtaining the well known ordinary X xave solution
(also called X-shaped pulse)

 \bb \Psi(\rho,\zeta) \; \equiv \; X \ug \frac{aV}{\sqrt{(aV-i\ze)^2 +
\left(\frac{V^2}{c^2}-1\right)\rho^2}}\label{S2X}\ee

--- Using spectrum (\ref{S2spec2}) and the identity (6.6444) of
ref.[19], we get

\bb \Psi(\rho,\zeta) \ug X \cdot J_0\left(\sqrt{\frac{V^2}{c^2}-1}
\;\; (aV)^{-2}d^2X^2\rho\right){\rm exp}\left[-(aV-i\zeta) \,
(aV)^{-2}d^2 \, X^2\right] \ee

--- The superluminal nondiffracting pulse

\bb \begin{array}{clcr} \Psi(\rho,\ze) \ug \sin^{-1} \, \left[
2\dis{\frac{d}{aV}}\left(\dis{ \sqrt{X^{-2}+(d/aV)^2+2\rho
d(aV)^{-2} \sqrt{V^2/c^2-1} } } \right.\right. \\

\\

 + \left.\left. \dis{ \sqrt{X^{-2}+(d/aV)^2-2\rho d(aV)^{-2}
 \sqrt{V^2/c^2-1}} } \;\right)^{-1} \right]\end{array} \ee
is obtained from spectrum (\ref{S2spec4}) using identity (6.752.1)
of ref.[19]  for $a>0$ and $d>0$.

From the previous discussion, we get to know that any solution obtained
from spectra of the type (\ref{S2specab}) with $\be_0=0$ is free
from noncausal (backwards) components.

In addition, \emph{when $\be_0=0$}, we can see that the pulsed
solutions depend on $z$ and $t$ through $\zeta=z-Vt$ only, and so
propagate rigidly, i.e. without distortion. Such pulses can be
transversally localized only if $V>c$, because if $V=c$ the
function $\Psi$ has to obey the Laplace equation on transverse
planes[12,13].

Many others superluminal localized waves can be easily
constructed[13] from the above solutions just by taking the
derivatives (of any order) with respect to $\zeta$. It is also
possible to show[13] that the new solutions obtained in this way
have their spectra shifted towards higher frequencies.

\

\emph{Now, let us consider, in eq.(\ref{S2geral5}), a spectrum of
the type (\ref{S2specab}) with $\be_0 > 0$:}

\bb A(\al,\be) \ug aV\delta(\be - \be_0)e^{-aV\al}
\label{S2specab2} \ee with $a$ a positive constant.

As we have seen, the presence of the delta function, with the
constant $\be_0>0$, implies that we are integrating (summing)
Bessel beams along the continuous line $\om = Vk_z + 2V\be_0$.
Now, the function $S(\al)=aV{\rm exp}(-aV\om)$ entails that we are
considering a frequency spectrum of the type $S(\om) \propto {\rm
exp}(-a\om)$, and therefore with a bandwidth given by $\Delta\om =
1/a$.

Since $\be_0>0$, the interval $V\be_0 \leq \om < 2V\be_0$ (or,
equivalently in this case, $0\leq \al < \be_0$), corresponds to
backward Bessel beams, i.e. negative values of $k_z$. However, we
can get physical solutions when making the contribution of this
frequency interval negligible. In this case, it can be done by
making $a\be_0 V<<1$, so that the exponential decay of the
spectrum $S$ with respect to $\om$ is very slow and the
contribution of the interval $\om\geq 2V\be_0$ (where $k_z\geq 0$)
overruns the $V\be_0 \leq \om < 2V\be_0$ (where $k_z<0$)
contribution.

Incidentally, we note that, once we ensure the causal behavior of
the pulse by making $aV\be_0<<1$ in (\ref{S2specab2}), we have
that $\Delta\al=1/aV>>\be_0$, and so we can simplify the
argument of the Bessel function, in the integrand of superposition
(\ref{S2geral5}), by neglecting the term $(V^2/c^2 -1)\be_0^2$.
With this, the superposition (\ref{S2geral5}), with the spectrum
(\ref{S2specab2}), can be written as

\bb \Psi(\rho,\zeta,\eta) \, \approx \, a\,V\,e^{-i\be_0\eta}\,
\int_{0}^{\infi} d\al
J_0\left(\rho\,\sqrt{\left(\frac{V^2}{c^2}-1\right)
 \al^2+2\left(\frac{V^2}{c^2}+1\right)\al\be_0}\, \right) \,
e^{i\al\ze}\,e^{-aV\al} \label{S2geral6}\ee

Now, we can use identity (6.616.1) of ref.[19] and obtain
the new localized superluminal solution called[13] Superluminal
Focus Wave Mode (SFWM):

\bb \Psi_{\rm SFWM}(\rho,\ze,\eta) \ug {\rm e}^{-i\be_0\eta} \; X
\; \dis{\exp\left[ \frac{\be_0(V^2+c^2)}{V^2-c^2} \, \left(
(aV-i\ze)- a\,VX^{-1} \right) \right]}\label{S2SFWM}\ee
 where, as before, $X$ is the ordinary X pulse (\ref{S2X}). The
 center of the SFWM is localized on $\rho=0$ and $\zeta=0\,$ (i.e. AT
 $z=Vt$). The intensity, $|\Psi|^2$, of this pulse propagates rigidly,
it being a
 function of $\rho$ and $\zeta$ only. However, the complex function
 $\Psi_{\rm SFWM}$ (i.e. its real and imaginary parts) propagate
 just with local variations, recovering their whole three
 dimensional form after each space and time interval given by $\Delta
z_0 =
 \pi/\be_0$ and $\Delta t_0 = \pi/\be_0 V$.

 The SFWM solution written above, for $V \longrightarrow c^+$
 reduces to the well known Focus Wave Mode (FWM) solution[11],
 travelling with speed $c$:

\bb \Psi_{\rm FWM}(\rho,\ze,\eta) \ug ac \dis{\frac{{\rm
e}^{-i\be_0\eta}}{ac-i\ze} \;
 \exp\left[-\frac{\be_0\rho^2}{ac-i\ze}\right]} \ . \ee

Let us also emphasize that, since $\be_0>0$, the spectrum
(\ref{S2specab2}) results to be constituted by angular frequenies
$\om \geq V\be_0$. Thus, our new solution can be used to construct
high frequency pulses.

\

\subsubsection{Finite energy nondiffracting pulses}

\

\

In this subsection, we will show how to get finite energy
localized wave pulses. These new waves can propagate for long
distances while maintaining their spatial resolution, i.e. they
possess a large depth of field.

As we have seen, ideal nondiffracting waves can be constructed by
superposing Bessel beams (eq.(\ref{S2geral4}) for cylindrical
symmetry) with a spectrum $A(\om,k_z)$ that satisfies a linear
relationship between $\om$ and $k_z$. In the general bidirectional
decomposition method, this can be obtained by using spectra of the
type (\ref{S2specab}) in superposition (\ref{S2geral5}).

Solutions of this type possess an infinity depth of field, however
they exhibit infinite energy[11,13]. To overcome this
problem, we can truncate an ideal nondiffracting wave by a finite
aperture, and the resulting pulse will have finite energy and a
finite field depth.  Even so, such field depths may be very large
when compared with those of ordinary waves.

The problem in this case is that the resulting field has to be
calculated from the diffraction integrals (such as the well known
Rayleigh-Sommerfeld formula) and, in general, a closed analytical
formula for the resulting pulse cannot be obtained.

However, there is another way to construct localized pulses with
finite energy[13]. That is, by using spectra $A(\om,k_z)$ in
(\ref{S2geral4}) whose domains are not restricted to be defined
exactly over the straight line $\om = Vk_z + b$, but around that
line, where the spectra should concentrate their main values. In
other words, the spectrum has to be well localized in the vicinity
of that line.

Similarly, in terms of the generalized bidirectional decomposition
given in (\ref{S2geral5}), finite energy nondiffracting wave
pulses can be constructed considering well localized spectral
functions $A(\al,\be)$ in the vicinity of the line $\be=\be_0$,
being $\be_0$ a constant.

To exemplify this method, let us consider the following spectrum

\bb A(\al,\be) \ug \left\{\begin{array}{clr}
 &a\,q\,V\,e^{-aV\al}e^{-q(\be-\be_0)}\ \ \ \ \ \ \ \ \ \ & {\rm for} \
 \be \geq \be_0  \\

 \\

 &0 \ \ \ \ \ \ \ \ \ \ \ \ \ \ \ \ \ \ & {\rm for} \
 0 \leq \be < \be_0
 \end{array} \right. \label{S2specSMPS}\ee
in the superposition (\ref{S2geral5}), quantities $a$ and $q$ being free
positive constants and $V$ the peak's pulse velocity (here $V\geq
c$).

It is easy to see that the above spectrum is zero in the region
above the $\be=\be_0$ line, while it decays in the region below
(as well as along) such a line. We can concentrate this spectrum
on $\be=\be_0$ by choosing values of $q$ in such a way that
$q\be_0>>1$. The faster the spectrum decay takes place in the
region below the $\be=\be_0$ line, the larger the field depth of
the corresponding pulse results to be.

Besides this, once we choose $q\be_0>>1$ to obtain pulses with
large field depth, we can also minimize the contribution of the
noncausal (backward) components by choosing $aV\be_0<<1$; in
analogy with the results we obtained for the SFWM case.

Still in analogy with the SFWM case, when we choose $q\be_0>>1$
(i.e. A long field depth) and $aV\be_0<<1$ (minimal contribution of
backward components), we can simplify the argument of the Bessel
function, in the integrand of superposition (\ref{S2geral5}), by
neglecting the term $(V^2/c^2 -1)\be_0^2$.

With the observations above, we can write the superposition
(\ref{S2geral5}) with the spectrum (\ref{S2specSMPS}) as

\bb \begin{array}{clcr} \Psi(\rho,\zeta,\eta) \, \approx & a\,q\,V
\dis{\int_{\be_0}^{\infi} d\be\,\int_{0}^{\infi} d\al\,
J_0\left(\rho\,\sqrt{\left(\frac{V^2}{c^2}-1\right)\al^2+2\left(\frac{V^2}{c^2}+1\right)\al\be}\,
\right)}\\

\\

& \times\,\dis{ e^{-i\be\eta}
e^{i\al\ze}e^{-q(\be-\be_0)}\,e^{-aV\al}}\end{array}
\label{S2intSMPS}\ee and, using identity (6.616.1) given in
ref.[19], we get

\bb \Psi(\rho,\zeta,\eta) \, \approx  q\,X \int_{\be_0}^{\infi}
d\be\,\dis{e^{-q(\be-\be_0)} e^{-i\be\eta}{\rm
exp}\left[\be\,\frac{V^2+c^2}{V^2-c^2}\left(aV-i\zeta -
aVX^{-1}\right) \right]}, \label{S2intSMPS2} \ee which can be
viewed as a superposition of the SFWM pulses (see
eq.(\ref{S2SFWM})).

The above integration can be easily made and results[13] in the so
called Superluminal Modified Power Spectrum (SMPS) pulse:

\bb \Psi_{\rm SMPS}(\rho,\ze,\eta) \ug q\,X \,\frac{{\rm
exp}[(Y-i\eta)\be_0]}{q-(Y-i\eta)} \label{S2SMPS}   \ee where $X$
is the ordinary X pulse (\ref{S2X}) and $Y$ is defined as

 \bb Y \ \equiv \ \dis{\frac{V^2+c^2}{V^2-c^2} \; \left((aV-i\ze)-
aVX^{-1} \right)} \ee

The SMPS pulse is a superluminal localized wave, with field
concentration around $\rho=0$ and $\zeta=0\,$ (i.e. in $z=Vt$),
and with finite total energy. We will show that the depth of
field, $Z$, of this pulse is given by $Z_{\rm SMPS} = q/2$.

An interesting property of the SMPS pulse is related with its
transverse width (the transverse spot size at the pulse center).
It can be shown from (\ref{S2SMPS}) that for the cases where
$aV<<1/\be_0$ and $q\be_0>>1$, i.e., for the cases considered by
us, the transverse spot size, $\Delta\rho$, of the pulse center
($\zeta=0$) is dictated by the exponential function in
(\ref{S2SMPS}) and is given by

\bb  \Delta\rho \ug c\,\sqrt{\frac{aV}{\be_0(V^2+c^2)} +
\frac{V^2-c^2}{4\be_0^2(V^2+c^2)^2}} \ee which clearly does not
depend on $z$, and so remains constant during the propagation. In
other words, in spite of the fact that the SMPS pulse suffers an
intensity decrease during the propagation, it preserves its
transverse spot size. This interesting characteristic is not
verified in ordinary pulses, like the gaussian ones, where the
amplitude of the pulse decreases and the width increases by the same
factor.

\begin{figure}[!h]
\begin{center}
 \scalebox{0.8}{\includegraphics{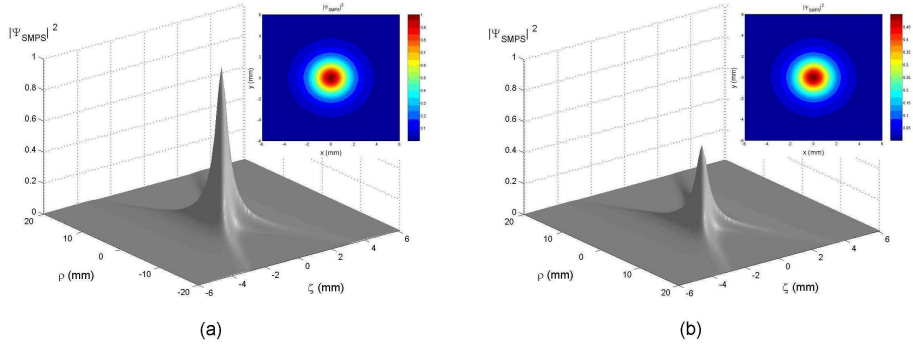}}
\end{center}
\caption{Representation of a Superluminal Modified Power Spectrum
pulse, eq.(\ref{S2SMPS}). \ Its total  energy is {\em finite}
(even without any truncation), and so it gets deformed
 while propagating, since its amplitude decreases with time. \ In
Fig.\ref{S2MHRFig1sec2}a we
 represent, for $t=0$, the pulse corresponding to $\be_0=33\,{\rm m}^{-1}$, $V=1.01 c$,
$a=10^{-12}\,$s and $q=10^5\,$m. \ In Fig.\ref{S2MHRFig1sec2}b it
is depicted the same pulse after having travelled $50 \; $km.}
\label{S2MHRFig1sec2}
\end{figure}

Figure \ref{S2MHRFig1sec2} shows a SMPS pulse intensity, with
$\be_0=33\,{\rm m}^{-1}$, $V=1.01 c$, $a=10^{-12}\,$s and
$q=10^5\,$m, at two different moments, for $t=0$ and after
$50\,$km of propagation, where, as we can see, the pulse becomes
less intense (half of its initial peak intensity). It can be noted
that, in spite of the intensity decrease, the pulse maintains its
transverse width, as one can see from the 2D plots in
Fig.(\ref{S2MHRFig1sec2}), which show the field intensities in the
transverse sections at $z=0$ and $z=q/2=50$ km.

\

Other three important well known \emph{finite energy}
nondiffracting solutions can be obtained directly from the SMPS
pulse:

--- The first one, obtained from (\ref{S2SMPS}) by making $\be_0=0$,
is the so called[13] Superluminal Splash Pulse (SSP),

\bb \Psi_{\rm SSP}(\rho,\ze,\eta) \ug \frac{q\,X}{q+i\eta-Y}
\label{S2SSP} \ee

--- The other two are luminal pulses. By taking the limit $V
\rightarrow c^+$ in the SMPS pulse (\ref{S2SMPS}), we get the well
known[11] \emph{luminal} Modified Power Spectrum (MPS) pulse

\bb \Psi_{\rm MPS}(\rho,\ze,\eta) \ug
\frac{a\,q\,c\,e^{-i\be_0\eta}}{(q+i\eta)(ac-i\zeta) +
\rho^2}\,{\rm exp}\left(\frac{-\be_0\rho^2}{ac-i\zeta}\right)
\label{S2MPS}\ee

Finally, by taking the limit $V \rightarrow c^+$ and making
$\be_0=0$ in the SMPS pulse (or, equivalently, by making $\be_0=0$
in the MPS pulse (\ref{S2MPS}), or,instead, by taking the limit $V
\rightarrow c^+$ in the SSP (\ref{S2SSP})), we obtain the well
known[11] \emph{luminal} Splash Pulse (SP) solution

\bb \Psi_{\rm SP}(\rho,\ze,\eta) \ug
\frac{a\,q\,c}{(q+i\eta)(ac-i\zeta) + \rho^2} \label{S2SP}\ee

It is also interesting to notice that the X and SFWM pulses can be
obtained from the SSP and SMPS pulses (respectively) by making $q
\rightarrow \infty$ in Eqs.(\ref{S2SSP}) and (\ref{S2SMPS}). As a
matter of fact, the solutions SSP and SMPS can be viewed as THE finite
energy versions of the X and SFWM pulses, respectively.

\

\emph{Some characteristics of the SMPS pulse:}

\

Let us examine the on-axis ($\rho=0$) behavior of the SMPS pulse.

On $\rho=0$ we have

\bb \Psi_{\rm SMPS}(\rho=0,\zeta,\eta) \ug aqV
e^{-i\be_0z}[(aV-i\zeta)(q+i\eta)]^{-1} \label{S2SMPSr0}\ee

From this expression, we can show that the longitudinal
localization $\Delta z$, for $t=0$, of the SMPS pulse square
magnitude is

\bb \Delta z \ug 2aV \ee

If we now define the field depth $Z$ as the distance over which
the pulse's peak intensity is $50\%$ at least of its initial
value\footnote{We can expect that while the pulse peak intensity
is maintained, also is its spatial form.}, then we can obtain,
from (\ref{S2SMPSr0}), the depth of field

\bb Z_{\rm SMPS} \ug \frac{q}{2} \label{S2ZSMPS} \ee which depends
only on $q$, as we expected since $q$ regulates the concentration
of the spectrum around the line $\om = Vk_z + 2V\be_0$.

Now, let us examine the maximum amplitude $M$ of the real part of
(\ref{S2SMPSr0}), which for $z=Vt$ writes ($\zeta=0$ and
$\eta=2z$):

\bb M_{\rm SMPS} \equiv {\rm Re}[\Psi_{\rm SMPS}(\rho=0,z=Vt)] \ug
\frac{\cos(2\be_0z) - 2(z/q)\sin(2\be_0z)}{1 + 4(z/q)^2}
\label{S2MSMPS}\ee

\

Initially, for $z=0$, $t=0$, one has $M=1$ and can also infer
that:

\

(i) when $z/q<<1$, namely, when $z<<Z$, equation (\ref{S2MSMPS})
becomes

\bb M_{\rm SMPS} \approx \cos(2\be_0 z)\;\;\;\;\;{\rm for}\;\;
z<<Z \ee and the pulse's peak actually oscillates harmonically with
``wavelength'' $\Delta z_0 = \pi / \be_0$ and ``period'' $\Delta
t_0 = \pi/V\be_0$, all along its field depth.


\

(ii) When $z/q >>1$, namely $z>>Z$, equation (\ref{S2MSMPS})
becomes

\bb M_{SMPS} \approx - \frac{\sin(2\be_0 z)}{2z/q} \;\;\;\;{\rm
for}\;\; z>>Z \ee

Therefore, beyond its depth of field, the pulse goes on oscillating
with the same $\Delta z_0$, but its maximum amplitude decays
proportionally to $z$.

In the next two Sections we are going to see an interesting
application of the localized wave pulses.

\section{Space-Time Focusing of X-shaped Pulses}\label{secIII}

In this Section we are going to show how one can in general use
any known Superluminal solution, to obtain from it a large number
of analytic expressions for space-time focused waves, endowed with
a very strong intensity peak at the desired location.

The method presented here is a natural extension of that developed
by A. Shaarawi et al.[20], where the space-time focusing was
achieved by superimposing a discrete number of ordinary X-waves,
characterized by different values $\theta$ of the axicon angle.

In this section, based on ref.[21], we will go on to more
efficient superpositions for varying velocities $V$, related to
$\theta$ through the known[3,4] relation $V=c/\cos\theta$. This
enhanced focusing scheme has the advantage of yielding analytic
(closed-form) expressions for the spatio-temporally focused
pulses.

Let us start considering an axially symmetric ideal nondiffracting
superluminal wave pulse $\psi(\rho ,z-Vt)$ in a dispersionless
medium, where $V = c/\cos\theta>c$ is the pulse velocity,
$\theta$ being the axicon angle. As we have seen in the previous
Section, pulses like these can be obtained by a suitable frequency
superposition of Bessel beams.

Suppose that we have now $N$ waves of the type
$\psi_n(\rho,z-Vn(t-tn))$, with different velocities
$c<V_1<V_2<..<V_N$, and emitted at (different) times $t_n$;
quantities $t_n$ being constants, while $n=1,2,...N$.  The center
of each pulse is located at  $z = V_n (t - t_n)$. To obtain a
highly focused wave, we need all the wave components
$\psi_n(\rho,z-Vn(t-tn))$ to reach the given point, $z = z_\frm$,
at the same time $t = t_\frm$. On choosing $t_1=0$ for the slowest
pulse $\psi_1$, it is easily seen that the peak of this pulse
reaches the point $z = z_\frm$ at the time $t_\frm = z_\frm /
V_1$. So we obtain that, for each $\psi_n$, the instant of emission
$t_n$ must be

\bb t_n \ug \left(\frac{1}{V_1}- \frac{1}{V_n}\right)z_\frm
\label{S3tn} \ee

With this, we can construct other exact solutions to the wave
equation, given by

\bb \Psi(\rho,z,t) \ug \int_{V_\minrm }^{V_\maxrm }\, \drm V \,
A(V) \, \psi\left(\rho,z-V\left(t-\left(\frac{1}{V_\minrm }-
\frac{1}{V}\right) z_\frm \right)\right) \label{S3int} \ , \ee
where $V$ is the velocity of the wave  $\psi(\rho,z-Vt)$ in the
integrand of (\ref{S3int}). In the integration, $V$ is considered
as a continuous variable in the interval $[V_\minrm ,V_\maxrm ]$.
In eq. (2), $A(V)$ is the velocity-distribution function that
specifies the contribution of each wave component (with velocity
$V$) to the integration. The resulting wave $\Psi(\rho,z,t)$ can
have a more or less strong amplitude peak at $z = z_\frm$, at time
$t_\frm = z_\frm / V_\minrm$, depending on $A(V)$ and on the
difference $V_\maxrm - V_\minrm$. Let us notice that also the
resulting wavefield will propagate with a Superluminal peak
velocity, depending on $A(V)$ too. In the cases when the
velocity-distribution function is well concentrated around a
certain velocity value, one can expect the wave (\ref{S3int}) to
increase its magnitude and spatial localization while propagating.
Finally, the pulse peak acquires its maximum amplitude and
localization at the chosen point $z = z_\frm$, and at time $t=
z_\frm / V_\minrm$, as we know. Afterwards, the wave suffers a
progressive spreading, and a decreasing of its amplitude.

\subsection{Focusing Effects by Using Ordinary X-Waves}

Here, we present a specific example by integrating (\ref{S3int})
over the standard, ordinary[4] X-waves, $X = aV[(aV-i(z-Vt))^2 +
(V^2/c^2 -1)\rho^2]^{-1/2}$. When using this ordinary X-wave, the
largest spectral amplitudes are obtained for low frequencies. For
this reason, one may expect that the solutions considered below
will be suitable mainly for low frequency applications.

Let us
choose, then, the function $\psi$ in the integrand of
eq.(\ref{S3int}) to be $\psi(\rho,z,t) \equiv X(\rho, z -
V(t-(1/V_\minrm -1/V)z_\frm))$, viz.

\bb \psi(\rho,z,t) \equiv X \ug
\dis{\frac{aV}{\sqrt{\left[aV-i\left( z - V\left( t -
\left(\frac{1} {V_\minrm }-\frac{1}{V}   \right)z_\frm \right)
\right)\right]^2 + \left(\frac{V^2}{c^2} -1 \right)\rho^2 }}}
\label{S3psiX} \ee

After some manipulations, one obtains the analytic {\em integral
solution}

\bb \Psi(\rho,z,t) \ug \int_{V_\minrm }^{V_\maxrm }\,
\dis{\frac{aV\,A(V)} {\sqrt{PV^2 + QV + R}} }\drm V \label{S3intx}
\ee with

\bb
\begin{array}{l}
P \ug \left[ \left( a+i\left(t-\frac{z_\frm}{V_\minrm
}\right)\right)^2
+ \frac{\rho^2}{c^2}\right]\\
\\
Q \ug 2\left(t-\frac{z_\frm}{V_\minrm } - ai\right)(z-z_\frm) \\
\\
R \ug \left[-(z-z_\frm)^2 - \rho^2 \right] \label{S3P}
\end{array}
\ee

\

In what follows, we illustrate the behavior of some new
spatio-temporally focused pulses, by taking into consideration
some different velocity distributions $A(V)$. These new pulses are
closed analytical \emph{exact} solutions of the wave equation.

\

\emph{First example:}

\

Let us consider our integral solution (\ref{S3intx}) with $A(V) =
1\,{\rm s/m}$. In this case, the contribution of the X-waves is
the same for all velocities in the allowed range $[V_\minrm
,V_\maxrm]$.

Using identity 2.264.2 listed in ref.[19], we get the particular
solution

\begin{figure}[!h]
\begin{center}
 \scalebox{1.1}{\includegraphics{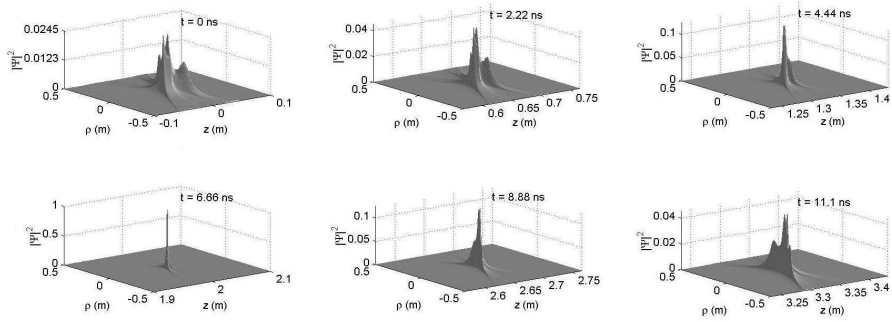}}
\end{center}
\caption{Space-time evolution of the Superluminal pulse
represented by eq.(\ref{S3solsec3}); the chosen parameter values
are $a=10^{-12}\;$s; \ $V_\minrm = 1.001 \; c$; \ $V_\maxrm =
1.005 \; c$ while the focusing point is at $z_\frm = 200\;$cm. \
One can see that this solution is associated with a rather good
spatio-temporal focusing. \ The field amplitude at $z = z_\frm$ is
40.82 times larger than the initial one. \ The field amplitude is
normalized at the space-time point $\rho = 0, \ z = z_\frm, \ t =
t_\frm$.} \label{S3MHRFig1sec3}
\end{figure}

\bb
\begin{array}{clcr}
\Psi(\rho,z,t) \!\!&=
\dis{\frac{a}{P}}\,\dis{\left(\sqrt{PV_\maxrm ^2 + QV_\maxrm + R}
- \sqrt{PV_\minrm ^2 + QV_\minrm  + R}\,\right)}\\
\\
&\;\;\; \dis{ + \frac{a\,Q}{2P^{3/2}}\,{\rm
ln}\left(\frac{2\,\sqrt{P(PV_\minrm ^2 + QV_\minrm  + R)} +
2PV_\minrm  + Q}{2\,\sqrt{P(PV_\maxrm ^2 + QV_\maxrm  + R)} +
2PV_\maxrm  + Q}\right)} \ , \label{S3solsec3}
\end{array}
\ee where $P$, $Q$ and $R$ are given in eq.(\ref{S3P}). A
3-dimensional (3D) plot of this function is provided in
Fig.\ref{S3MHRFig1sec3}; where we have chosen $a=10^{-12}$ s, \
$V_\minrm =1.001\; c$, \ $V_\maxrm =1.005\; c$ and $z_\frm
=200\;$cm. It can be seen that this solution exhibits a rather
evident space-time focusing. An initially spread-out pulse (shown
for $t=0$) becomes highly localized at $t=t_\frm=z_\frm
/V_\minrm=6.66\;$ns, the pulse peak amplitude at $z_\frm$ being
$40.82$ times greater than the initial one. In addition, at the
focusing time $t_\frm$ the field is much more localized than at
any other times. The velocity of this pulse is approximately
$V=1.003\; c$.

\

\

\emph{Second example:}

\

In this case we choose $A(V) = 1/V\,\,({\rm s/m})$, and, using the
identity 2.261 in ref.[19], eq.(\ref{S3intx}) gives

\bb \Psi(\rho,z,t) \ug \dis{\frac{a}{\sqrt{P}}\,{\rm ln}
\left(\frac{2\,\sqrt{P(PV_\maxrm ^2 + QV_\maxrm  + R)} +
2PV_\maxrm  + Q} {2\,\sqrt{P(PV_\minrm ^2 + QV_\minrm  + R)} +
2PV_\minrm  + Q} \right)} \ . \label{S3dois} \ee

\

Other exact closed-form solutions can be obtained[21] considering, for
instance, velocity distributions like $A(V)=1/V^2$ and $A(V)=1/V^3$.

Actually, we can construct many others spatio-temporally focused
pulses from the above solutions, just by taking their time derivatives
(of any order). It is also possible to show[21] that the new
solutions obtained in this way have their spectra shifted towards
higher frequencies.

\section{Chirped Optical X-Type Pulses in Material Media}\label{secIV}

The theory of the localized waves was initially developed for free
space (vacuum). In 1996, S\~{o}najalg et al.[5] showed that the
localized wave theory can be extended to include (unbounded)
dispersive media. This was obtained by making the axicon angle of
the Bessel beams (BBs) vary with the frequency[5-7] in such a way
that a suitable frequency superposition of these beams does
compensate for the material dispersion. Soon after this idea was
reported, many interesting nondiffracting/nondispersive pulses
were obtained theoretically[5-7] and experimentally[5].

In spite of this extended method to be of remarkable importance,
working well in theory, its experimental implementation is not so
simple\footnote{We refer the interested reader to quotations
[5-7] for obtaining a description, theoretical and experimental,
of this extended method}.

In 2004 Zamboni-Rached et al.[22] developed a simpler way to
obtain pulses capable of recovering their spatial shape, both
transversally and longitudinally, after some propagation. It
consisted in using chirped optical X-typed pulses, while keeping
the axicon angle fixed. Let us recall that, by contrast, chirped
Gaussian pulses in unbounded material media may recover only their
longitudinal shape, since they undergo a progressive transverse
spreading while propagating.

The present section is devoted to this approach.


Let us start with an axis-symmetric Bessel beam in a material
medium with refractive index $n(\om)$:

\bb \psi(\rho,z,t)\ug J_0(k_{\rho}\rho) \, \exp(i\be z) \,
\exp(-i\om t) \; ,
\label{S4bb}\ee  
where it must be obeyed the condition \
$k_{\rho}^2=n^2(\om)\om^2/c^2 - \be^2$, \ which connects among
themselves the transverse and longitudinal wave numbers $k_{\rho}$
and $\be$, and the angular frequency $\om$.  In addition, we
impose that $k_{\rho}^2 \geq 0$ and $\om / \be \geq 0$, to avoid a
nonphysical behavior of the Bessel function $J_0(.)$ and to
confine ourselves to forward propagation only.

Once the conditions above are satisfied, we have the liberty of
writing the longitudinal wave number as $\be = (n(\om)\om
\cos\theta)/c$ and, therefore, $k_{\rho} = (n(\om)\om
\sin\theta)/c$; where (as in the free space case) $\theta$ is the
axicon angle of the Bessel beam.

Now we can obtain a X-shaped pulse by performing a frequency
superposition of these Bessel beams [BB], with $\be$ and $k_{\rho}$ given by the
previous relations:

\bb \Psi(\rho,z,t) \ug \int_{-\infi}^{\infi}\,
S(\om)\,J_0\left(\frac{n(\om)\om}{c} \sin\theta\,\rho\right)\,
\exp[i\be(\om)z]\, \exp(-i\om t)\,\drm\om \; ,
\label{S4geral}\ee        
\\
where $S(\om)$ is the frequency spectrum, and the axicon angle is
kept constant.

One can see that the phase velocity of each BB in our
superposition (\ref{S4geral}) is different, and given by $V_{\rm
phase} = c/(n(\om)\cos\theta)$. So, the pulse represented by
eq.(\ref{S4geral}) will suffer dispersion during its propagation.

As we said, the method developed by S\~onajalg et al.[5] and
explored by others[6,7], to overcome this problem, consisted in
regarding the axicon angle $\theta$ as a function of the
frequency, in order to obtain a linear relationship between $\be$
and $\om$.

Here, however, we wish to work with a {\em fixed} axicon angle,
and we have to find out another way for avoiding dispersion and
diffraction along a certain propagation distance. To do that, we
might choose a chirped gaussian spectrum  $S(\om)$ in
eq.(\ref{S4geral})

\bb S(\om) \ug
\frac{T_0}{\sqrt{2\pi(1+iC)}}\,\,\exp[-q^2(\om-\om_0)^2]
\label{S4S}\;\;\;\;{\rm with}\;\;\;\;\; q^2 \ug
\frac{T_0^2}{2(1+iC)} \; , \label{S4s}\ee           
where $\om_0$ is the central frequency of the spectrum, $T_0$ is
a constant related with the initial temporal width,  and $C$ is
the chirp parameter (we chose as temporal width the half-width of
the relevant gaussian curve when its heigth equals $1/e$ times its
full heigth).  Unfortunately, there is no analytical solution to
eq.(\ref{S4geral}) with $S(\om)$ given by eq.(\ref{S4s}), so that
some approximations are to be made.

Then, let us assume  that the spectrum $S(\om)$, in the
surrounding of the carrier frequency $\om_0$ , is enough narrow
that $\Delta\om/\om_0<<1$, so to ensure that $\be(\om)$ can be
approximated by the first three terms of its Taylor expansion in
the vicinity of $\om_0$: That is, \ $\be(\om)\approx \be(\om_0) +
\be'(\om)|_{\om_0}\,(\om - \om_0) + (1/2) \be''(\om)|_{\om_ 0}\,
(\om - \om_0)^2$; \ where, after using $\be =n(\om)\om
\cos\theta/c$, it results that

\bb \frac{\pa \be}{\pa\om} \ug \frac{\cos\theta}{c}\left[n(\om) +
\om\,\frac{\pa n}{\pa\om} \right]\,; \ \ \ \frac{\pa^2
\be}{\pa\om^2} \ug \frac{\cos\theta}{c}\left[ 2\frac{\pa
n}{\pa\om} + \om \frac{\pa^2 n}{\pa\om^2} \right] \; .
\label{S4b1}\ee

\

As we know, $\be'(\om)$  is related to the pulse group-velocity by
the relation $Vg = 1/ \be'(\om)$.  Here we can see the difference
between the group-velocity of the X-type pulse (with a fixed
axicon angle) and that of a standard gaussian pulse.  Such a
difference is due to the factor $\cos\theta$ in eq.(\ref{S4b1}).
Because of it, the group-velocity of our X-type pulse is always
greater than the gaussian's. In other words, $(V_\grm)_\X =
(1/\cos\theta)(V_\grm)_{gauss}$.

We also know that the second derivative of $\be(\om)$ is related
to the group-velocity dispersion (GVD) $\be_2$ by $\be_2 =
\be''(\om)$.

The GVD is responsible for the temporal (longitudinal) spreading
of the pulse. Here one can see that the GVD of the X-type pulse is
always smaller than that of the standard gaussian pulses, due the
factor $\cos\theta$ in eq.(\ref{S4b1}).  Namely: $(\be_{2})_\X =
\cos\theta (\be_{2})_{\rm gauss}$.

Using the above results, we can write

\

\bb \begin{array}{clcr}
 \Psi(\rho,z,t) \!\!&= \dis{\frac{T_0\,\,\exp[i\be(\om_0)z]\,
\exp(-i\om_0 t)}{\sqrt{2\pi(1+iC)}}
\,\int_{-\infi}^{\infi}\drm\om\,J_0\left(\frac{n(\om)\om}{c}
\sin\theta\,\rho\right)}\\
\\
&\;\;\;\times\,\dis{\exp\left\{i\frac{(\om-\om_0)}{V_\grm}\left[z
- V_\grm t \right] \right\} \,
\exp\left\{(\om-\om_0)^2\left[\frac{i\be_2}{2}z - q^2
\right] \right\}} \; . \label{S4geral2} \end{array} \ee  

\

The integral in eq.(\ref{S4geral2}) cannot be solved analytically,
but it is enough for us to obtain the pulse behavior. Let us
analyze the pulse for $\rho=0$. In this case we obtain:

\bb \Psi(\rho=0,z,t) \ug \dis{\frac{T_0\,\exp[i\be(\om_0)z]\,
\exp(-i\om_0 t)}{\sqrt{T_0^2 -
i\be_2(1+iC)z}}\,\exp\left[\frac{-(z-V_\grm
t)^2(1+iC)}{2V_\grm^2[T_0^2 - i\be_2(1+iC)z]}\right]} \; .
\label{S4sol1} \ee

From eq.(\ref{S4sol1}) one can immediately see that the initial
temporal width of the pulse intensity is $T_0$ and that, after
a propagation distance $z$, the time-width $T_1$ becomes

\bb \frac{T_1}{T_0} \ug \left[\left(1+\frac{C\be_2z}{T_0^2}
\right)^2 + \left(\frac{\be_2z}{T_0^2}\right)^2\right]^{1/2} \; .
\label{S4T1}\ee

Relation (\ref{S4T1}) describes the pulse spreading-behavior. One
can easily show that such a  behavior depends on the sign
(positive or negative) of the product $\be_2 C$, as is well known
for the standard gaussian pulses[23].

In the case $\be_2C > 0$,  the pulse will monotonically become
broader and broader with the distance $z$.  On the other hand, if
$\be_2C < 0$ the pulse will suffer, in a first stage, a narrowing,
and then it will spread during the rest of its propagation. So,
there will be a certain propagation distance AT which the pulse
will recover its initial temporal width ($T_1=T_0$). From relation
(\ref{S4T1}), we can find this distance $Z_{T1=T_0}$ (considering
$\be_2 C < 0$) to be

\bb Z_{T_1=T_0} \ug \frac{-2CT_0^2}{\be_2(C^2+1)} \; . \label{S4Z}
\ee

One may notice that the maximum distance at which our chirped
pulse, with given $T_0$ and $\be_2$,  may recover its initial
temporal width can  be easily evaluated from eq.(\ref{S4Z}), and
it results to be $L_{\rm disp} = T_0^2 / \be_2$.  We shall call
such a maximum value $L_{\rm disp}$ the ``dispersion length". It
is the maximum distance the X-type pulse may travel while
recovering its initial longitudinal shape. Obviously, if we want
the pulse to reassume its longitudinal shape at some desired
distance $z < L_{\rm disp}$, we have just to suitably choose the
value of the chirp parameter.

Let us emphasize that the property of recovering its own initial
temporal (or longitudinal) width may be verified to exist also in
the case of chirped standard gaussian pulses. However, the latter
will suffer a progressive transverse spreading, which will not be
reversible. The distance at which a gaussian pulse doubles its
initial transverse width $w_0$ is $z_{\rm diff} = \sqrt{3}\pi
w_0^2/\lambda_0$, where $\lambda_0$ is the carrier wavelength.
Thus, we can see that optical gaussian pulses with great
transverse localization will get spoiled in a few centimeters or
even less.

Now we shall show that it is possible to recover also the
transverse shape of  the chirped X-type pulse intensity; actually,
it is possible to recover its entire spatial shape after a
distance $Z_{T_1=T_0}$.

To see this, let us go back to our integral solution
(\ref{S4geral2}), and perform the change of coordinates \ $(z,t)
\rightarrow (\Delta z, t_c = z_c/V_\grm)$, with
\bb\left\{\begin{array}{l} z \ug z_c + \Delta z\\
\\
 \dis{t=t_c\equiv \frac{z_c}{V_\grm}} \end{array}\right. \label{S4zc}\ee
where $z_c$ is the center of the pulse ($\Delta z$ is the distance
from such a point), and $t_c$ is the time at which the pulse
center is located at $z_c$. What we are going to do is comparing
our integral solution (\ref{S4geral2}), when $z_c=0$ (initial
pulse), with that when $z_c=Z_{T_1=T_0}=-2CT_0^2/(\be_2(C_2+1))$.

In this way, the solution (\ref{S4geral2}) can be written, when
$z_c = 0$, as

\bb \begin{array}{clcr} \Psi(\rho,z_c=0,\Delta z)
\!\!&=\dis{\frac{T_0\,\,\, \exp(i\be_0\Delta
z)}{\sqrt{2\pi(1+iC)}}}\,\int_{-\infi}^{\infi}\drm\om\,
J_0(k_{\rho}(\om)\rho)\,\exp\left[\frac{-T_0^2\,(\om-\om_0)^2}{2(1+C^2)}\right]\\
\\
\\ &\;\;\;\dis{\times\,\exp\left\{i\left[\frac{(\om-\om_0)\Delta
z}{V_\grm} + \frac{(\om-\om_0)^2\be_2\Delta z}{2} +
\frac{(\om-\om_0)^2T_0^2C}{2(1+C^2)} \right]
\right\}}\\
\ \label{S4zc0} \end{array}\ee      
where we have taken the value $q$ given by (\ref{S4s}).

To verify that the pulse intensity recovers its entire original
form at $z_c = Z_{T_1=T_0} = -2\,CT_0^2/ [\be_2(C^2+1)]$, we can
analyze our integral solution at that point, obtaining

\

\bb \begin{array}{l} \Psi(\rho,z_c=Z_{T_1=T_0},\Delta z) \ugg
\dis{\frac{T_0\,\,\dis{\exp\left\{i\be_0 \left[z_c-\Delta z'-
\frac{cz_c}{\cos\theta\,n(\om_0)V_\grm}\right]\right\}}}{\sqrt{2\pi(1+iC)}}}\\
\\
\\
\times \dis{\int_{-\infi}^{\infi}} \dis{
\drm\om\,J_0(k_{\rho}(\om)\rho)\,\exp\left[\frac{-T_0^2\,(\om-\om_0)^2}{2(1+C^2)}
\right]}\\
\\
\\
\dis{\times\,\exp\left\{-i\left[\frac{(\om-\om_0)\Delta
z'}{V_\grm} + \frac{(\om-\om_0)^2\be_2\Delta z'}{2} +
\frac{(\om-\om_0)^2T_0^2C}{2(1+C^2)} \right]
\right\}}\\
\label{S4zt1}\end{array}\ee               
\\
where we put  $\Delta z = -\Delta z'$. \ In this way, one
immediately sees that

\

\bb |\Psi(\rho,z_c=0,\Delta z)|^2 =
|\Psi(\rho,z_c=Z_{T_1=T_0},-\Delta z)|^2 \; . \label{S4inten}
\ee                                        

\

Therefore, from eq.(\ref{S4inten}) it is clear that the chirped
optical X-type pulse intensity will recover its original
three-dimensional form, with just a longitudinal inversion at the
pulse center: The present method being, in this way, a simple and
effective procedure for compensating the effects of diffraction
and dispersion in an unbounded material medium; and a method
simpler than the one of varying the axicon angle with the
frequency.

Let us stress that we can choose the distance $z = Z_{T_1=T_0}\leq
L_{\rm disp}$ at which the pulse will take on again its spatial
shape by choosing a suitable value of the chirp parameter.

Till now, we have shown that the chirped X-type pulse recovers its
three-dimensional shape after some distance, and we have also
obtained an analytic description of the pulse {\em longitudinal}
behavior (for $\rho=0$) during propagation, by means of
eq.(\ref{S4sol1}). However, one does not get the same information
about the pulse transverse behavior: We just know that it is
recovered at $z=Z_{T_1=T_0}$.

So, to complete the picture, it would be interesting if we could
get also the {\em transverse} behavior in the plane of the pulse
center $z=V_\grm t$.  In that way, we would obtain quantitative
information about the evolution of the pulse-shape during its
entire propagation.

We will not examine the mathematical details here; we just affirm
that THE transverse behavior of the pulse (in the plane $z=z_c=V_\grm
t$) during its whole propagation can approximately be described by

\bb \begin{array}{l} \Psi(\rho,z=z_c,t=z_c/V_\grm) \, \approx \,
 \dis{\frac{T_0\,\,\, \exp[i\be(\om_0)z]\, \exp(-i\om_0
t)}{\sqrt{2\pi(1+iC)}}\,\,\frac{\exp\dis{\left[\frac{-\tan^2\theta\,\rho^2}
{8\,V_\grm^2(-i\be_2z_c/2 \, + q^2)}
\right]}}{\sqrt{-i\be_2 z_c/2 \, + q^2}}}\\
\\
\\
\times\,\dis{\left[\Gamma(1/2)J_0\left(\frac{n(\om_0)\,\om_0\,\sin\theta\,\rho}
{c}\right)I_0\left(
\frac{\tan^2\theta\,\rho^2}{8\,V_\grm^2(-i\be_2 z_c/2 \, +
q^2)}\right)
\right.}\\
\\
\\
\left. + \dis{
2\sum_{p=1}^{\infty}\frac{2^p\Gamma(p+1/2)\Gamma(p+1)}{\Gamma(2p+1)}\,\,
J_{2p}\left(\frac{n(\om_0)\,\om_0\,\sin\theta\,\rho}{c}\right)I_{2p}\left(
\frac{\tan^2\theta\,\rho^2}{8\,V_\grm^2(-i\be_2 z_c/2 \, +
q^2)}\right)} \right]
\\
\ \label{S4trans1}
\end{array}\ee         
\\
where $I_p(.)$ is the modified Bessel function of the first kind
of order $p$, quantity $\Gamma(.)$ being the gamma function and
$q$ given by (\ref{S4s}).

The interested reader can consult ref.[22] for details on how to
obtain eq.(\ref{S4trans1}) from eq.(\ref{S4geral2}).

At a first sight, this solution could appear very complicated, but
{\em the series in its r.h.s. gives a negligible
contribution}. This fact renders our solution (\ref{S4trans1}) of
important practical interest and we will use it in the following.

For additional information about the transverse pulse evolution
(extracted from eq.(\ref{S4trans1})), the reader can consult again
ref.[22] . In the same reference, it is analyzed the effect
of a finite aperture generation on the chirped X-type pulses.

\

\subsection{An example: Chirped Optical X-typed pulse in bulk fused
Silica}

\

For a bulk fused Silica, the refractive index $n(\om)$ can be
approximated by the Sellmeier equation[23]

\

\bb n^2(\om) \ug
\dis{1\,+\,\sum_{j=1}^{N}\,\frac{B_j\,\om_j^2}{\om_j^2- \om^2}} \;
, \label{S4Selm}\ee \ where $\om_j$ are the resonance frequencies,
$B_j$ the strength of the $j$th resonance, and $N$ the total
number of the material resonances that appear in the frequency
range of interest. For our purposes it is appropriate to choose
$N=3$, which yields, for bulk fused silica[23], the values \
$B_1=0.6961663$; \ $B_2=0.4079426$; \ $B_3=0.8974794$; \
$\lambda_1=0.0684043 \; \mu \;$m; \ $\lambda_2=0.1162414 \;
\mu\;$m; \ and $\lambda_3=9.896161 \; \mu\;$m.

Now, let us consider in this medium a chirped X-type pulse, with
$\lambda_0=0.2\,\mu$m, \ $T_0=0.4 \;$ps, \ $C=-1$ and with an
axicon angle $\theta=0.00084 \;$rad, which correspond to an
initial central spot with $\Delta\rho_0=0.117 \;$mm.

\begin{figure}[!h]
\begin{center}
\scalebox{1.4}{\includegraphics{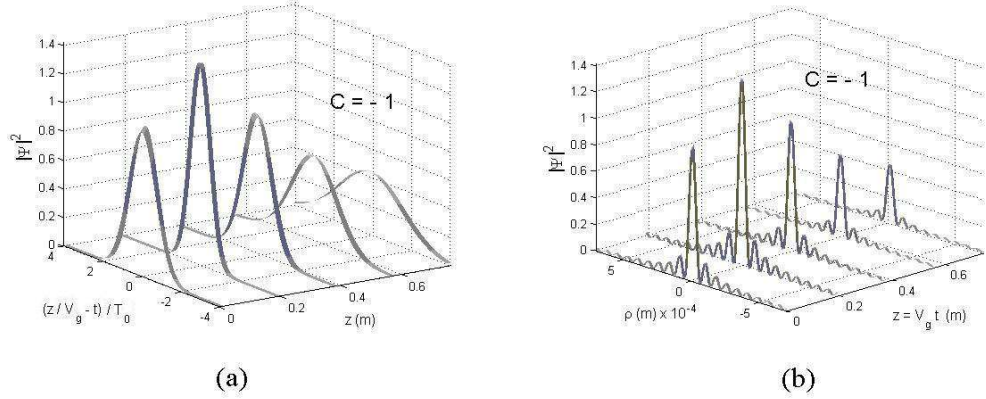}}
\end{center}
\caption{(a): Longitudinal-shape evolution of a chirped X-type
pulse, propagating in fused silica with $\lambda_0=0.2\mu \;$m, \
$T_0=0.4 \;$ps, \ $C=-1$ and axicon angle $\theta=0.00084 \;$rad,
which correspond to an initial transverse width of
$\Delta\rho_0=0.117 \;$mm. \ \ (b): Transverse-shape evolution for
the same pulse.} \label{S4MHRFig4sec4}
\end{figure}

We get from eqs. (\ref{S4sol1}) and (\ref{S4trans1}), the
longitudinal and transverse pulse evolution, which are represented
by Fig.\ref{S4MHRFig4sec4}

From Fig.\ref{S4MHRFig4sec4}(a), we can notice that, initially,
the pulse suffers a longitudinal narrowing with an increase of
intensity till the position $z=T_0^2/2\be_2=0.186$m. After this
point, the pulse starts to broaden decreasing its intensity and
recovering its entire longitudinal shape (width and intensity) at
the point $z=T_0^2/\be_2=0.373$m, as it was predicted.

At the same time, from Fig.\ref{S4MHRFig4sec4}(b), one can notice that
the pulse maintains its transverse width
$\Delta\rho=2.4\,c/(n(\om_0)\om_0 \sin\theta)=0.117 \;$mm \
(because $T_0\om_0>>1$) during its entire propagation; however,
the same does not occur with the pulse intensity. Initially the
pulse suffers an increase of intensity till the position
$z_c=T_0^2/2\be_2=0.186$m; after this point, the intensity starts
to decrease, and the pulse recovers its entire transverse shape at
the point $z_c=T_0^2/\be_2=0.373$m, as it was expected by us. Here
we have skipped the series on the r.h.s. of
eq.(\ref{S4trans1}), because, as we already said, it yields a
negligible contribution.

Summarizing, from Fig.\ref{S4MHRFig4sec4}, we can see that the
chirped X-type pulse recovers totally its longitudinal and
transverse shape at the position $z = L_{\rm Disp} =
T^2_0/\be_2=0.373 \;$m, as we expected.

Let us recall that a \emph{chirped gaussian pulse} may just
recover its longitudinal width, but with an intensity decrease, at
the position given by $z = Z_{T_1=T_0}=L_{\rm disp} =
T^2_0/\be_2$. Its transverse width, on the other hand, suffers a
progressive and irreversible spreading.

\

\section{Modeling the Shape of
Stationary Wave Fields: Frozen Waves}

In this Section we develop a very simple method[10,16,17], by having
recourse to superpositions of forward propagating and
\emph{equal-frequency} Bessel beams, that allows one controlling
the {\em longitudinal} beam-intensity shape within a chosen
interval $0\leq z \leq L$, where $z$ is the propagation axis and
$L$ can be much greater than the wavelength $\lambda$ of the
monochromatic light (or sound) which is being used. Inside such a
space interval, indeed, we succeed in constructing a {\em
stationary} envelope whose longitudinal intensity pattern can
approximately assume any desired shape, including, for instance,
one or more high-intensity peaks (with distances between them much
larger than $\lambda$); and which results ---in addition--- to be
naturally endowed also with a good transverse localization. Since
the intensity envelopes remains static, i.e. with velocity $V=0$,
we call ``Frozen Waves" (FW) such new solutions[10,16,17] to the
wave equations.

Although we are dealing here with exact solutions of the scalar
wave equation, vectorial solutions of the same kind for the
electromagnetic field can be obtained, since solutions to
Maxwell's equations follow naturally from the scalar wave equation
solutions[24,25].

First, we present the method considering lossless media[16,17]
and, in the second part of this section, we extend the method to
absorbing media[10].

\subsection{Stationary wavefields
with arbitrary longitudinal shape in lossless media, obtained by
superposing equal-frequency Bessel beams}

Let us start from the well-known axis-symmetric zeroth order
Bessel beam solution to the wave equation:

\bb \psi(\rho,z,t)\ug J_0(k_{\rho}\rho)e^{i\be z}e^{-i\om t}
\label{S5bb}\ee with

\bb k_{\rho}^2=\frac{\om^2}{c^2} - \be^2 \; , \label{S5k}  \ee
where $\om$, $k_{\rho}$ and $\be$ are the angular frequency, the
transverse and the longitudinal wave numbers, respectively. We
also impose the conditions

\bb \om/\be > 0 \;\;\; {\rm and}\;\;\; k_{\rho}^2\geq 0
\label{S5c2} \ee (which imply $\om/\be \geq c$) to ensure forward
propagation only (with no evanescent waves), as well as a physical
behavior of the Bessel function $J_0$.

Now, let us make a superposition of $2N + 1$ Bessel beams with the
same frequency $\om_0$, but with {\em different} (and still
unknown) longitudinal wave numbers $\be_m$:

\bb \dis{\Psi(\rho,z,t) \ug e^{-i\,\om_0\,t}\,\sum_{m=-N}^{N}
A_m\,J_0(k_{\rho\,m}\rho)\,e^{i\,\be_m\,z} } \; , \label{S5soma}
\ee where thye $m$ represent integer numbers and the $A_m$ are constant
coefficients. For each $m$, the parameters $\om_0$, $k_{\rho\,m}$
and $\beta_m$ must satisfy (\ref{S5k}), and, because of conditions
(\ref{S5c2}), when considering $\om_0 > 0$, we must have

\bb 0 \leq \be_m \leq \frac{\om_0}{c} \; . \label{S5be} \ee

Let us now suppose that we wish $|\Psi(\rho,z,t)|^2$, given by
eq.(\ref{S5soma}), to assume on the axis $\rho=0$ the pattern
represented by a function $|F(z)|^2$, inside the chosen interval
$0 \leq z \leq L$. In this case, the function $F(z)$ can be
expanded, as usual, in a Fourier series:

\

$$F(z) \ug
\sum_{m=-\infty}^{\infty}\,B_m\,e^{i\,\frac{2\pi}{L}\,m\,z} \; ,$$

\ where

\

$$B_m \ug \frac{1}{L} \dis{
\int_{0}^{L}\,F(z)\,e^{-i\,\frac{2\pi}{L}\,m\,z}\,\drm\,z } \ .$$

\

More precisely, our goal is finding out, now, the values of the
longitudinal wave numbers $\be_m$ and the coefficients $A_m$ of
(\ref{S5soma}), in order to reproduce approximately, within the
said interval $0 \leq z \leq L$ (for $\rho=0$), the predetermined
longitudinal intensity-pattern $|F(z)|^2$. \ Namely, we wish to
have

\bb \left|\,\sum_{m=-N}^{N} A_m e^{i\,\be_m\,z}\right|^{\,2}
\approx |F(z)|^{\,2} \;\;\;\; {\rm with}\;\;\; 0\leq z \leq L  \;
. \label{S5soma1} \ee

\

Looking at eq.(\ref{S5soma1}), one might be tempted to take $\be_m
= 2\pi m/L$, thus obtaining a truncated Fourier series, expected
to represent approximately the desired pattern $F(z)$. \
Superpositions of Bessel beams with $\be_m = 2\pi m/L$ have been
actually used in some works to obtain a large set of {\it
transverse} amplitude profiles[26]. However, for our purposes,
this choice is not appropriate, due to two principal reasons: \ 1)
It yields negative values for $\be_m$ (when $m<0$), which implies
backwards propagating components (since $\om_0 > 0$); \ 2) In the
cases when $L>>\lambda_0$, which are of our interest here, the
main terms of the series would correspond to very small values of
$\be_m$, which results in a very short field-depth of the
corresponding Bessel beams (when generated by finite apertures),
preventing the creation of the desired envelopes far form the
source.

Therefore, we need to make a better choice for the values of
$\be_m$, which permits forward propagation components only, and a
good depth of field. \ This problem can be solved by putting

\bb \be_m \ug Q + \frac{2\,\pi}{L}\,m \; , \label{S5be2}  \ee
where $Q>0$ is a value to be chosen (as we shall see) according to
the given experimental situation, and the desired degree of {\em
transverse} field localization. \ Due to eq.(\ref{S5be}), we get

\bb 0\leq Q \pm \frac{2\,\pi}{L}\,N \leq \frac{\om_0}{c} \; .
\label{S5N} \ee

Inequality (\ref{S5N}), can be used to determine the maximum value
of $m$, that we call $N_{\rm max}$, once $Q$, $L$ and $\om_0$ have
been chosen.

As a consequence, for getting a longitudinal intensity pattern
approximately equal to the desired one, $|F(z)|^2$, in the
interval $0\leq z \leq L $, eq.(\ref{S5soma}) should be rewritten
as

\bb \dis{\Psi(\rho=0,z,t) \ug
e^{-i\,\om_0\,t}\,e^{i\,Q\,z}\,\sum_{m=-N}^{N}
A_m\,e^{i\,\frac{2\pi}{L}m\,z} } \; , \label{S5soma2} \ee with

\bb A_m \ug \frac{1}{L} \dis{
\int_{0}^{L}\,F(z)\,e^{-i\,\frac{2\pi}{L}\,m\,z}\,\drm\,z } \; .
\label{S5An} \ee

Obviously, one obtains only an approximation to the desired
longitudinal pattern, because the trigonometric series
(\ref{S5soma2}) is necessarily truncated ($N \leq N_{\rm max}$).
Its total number of terms, let us repeat, will be fixed once the
values of $Q$, $L$ and $\om_0$ are chosen.

When $\rho \neq 0$, the wavefield $\Psi(\rho,z,t)$ becomes

\bb \dis{\Psi(\rho,z,t) \ug
e^{-i\,\om_0\,t}\,e^{i\,Q\,z}\,\sum_{m=-N}^{N}
A_m\,J_0(k_{\rho\,m}\,\rho)\,e^{i\,\frac{2\pi}{L}m\,z} } \; ,
\label{S5soma3} \ee with

\bb k_{\rho\,m}^2 \ug \om_0^2 - \left(Q + \frac{2\pi\,m}{L}
\right)^2 \; . \label{S5krn} \ee

The coefficients $A_m$ will yield {\em the amplitudes} and {\em
the relative phases} of each Bessel beam in the superposition.

Because we are adding together zero-order Bessel functions, we can
expect a {\em high} field concentration around $\rho=0$. Moreover,
due to the known non-diffractive behavior of the Bessel beams, we
expect that the resulting wavefield will preserve its transverse
pattern in the entire interval $0\leq z \leq L $.

The methodology developed here deals with the longitudinal
intensity pattern control. Obviously, we cannot get a total 3D
control, due the fact that the field must obey the wave equation.
However, we can use two ways to have some control over the
transverse behavior too. The first is through the parameter $Q$ of
eq.(\ref{S5be2}). \ Actually, we have some freedom in the choice
of this parameter, and FWs representing the same longitudinal
intensity pattern can possess different values of $Q$. The
important point is that, in superposition (\ref{S5soma3}), using a
smaller value of $Q$ makes the Bessel beams possess a higher
transverse concentration (because, on decreasing the value of $Q$,
one increases the value of the Bessel beams transverse wave
numbers), and this will reflect in the resulting field, which will
present a narrower central transverse spot. The second way to
control the transverse intensity pattern is using higher order
Bessel beams, and we shall show this in Section 5.1.1.

Now, let us present a few examples of our methodology.

\

\emph{First example:}

\

Let us suppose that we wish an optical wavefield with $\lambda_0 =
0.632\;\mu$m, i.e. with $\om_0 = 2.98 \times 10^{15}\;$Hz, whose
longitudinal pattern (along its $z$-axis) in the range $0 \leq z
\leq L$ is given by the function

 \bb
 F(z) \ug \left\{\begin{array}{clr}
 -4\,\,\dis{\frac{(z-l_1)(z-l_2)}{(l_2 - l_1)^2}} \;\;\; & {\rm
for}\;\;\; l_1 \leq z \leq l_2  \\
\\
 \;\;\;\;\;\;\;\;1 & {\rm for}\;\;\; l_3 \leq z \leq l_4 \\
\\
 -4\,\,\dis{\frac{(z-l_5)(z-l_6)}{(l_6 - l_5)^2}} & {\rm for}\;\;\; l_5
\leq z \leq
 l_6 \\
 \\
 \;\;\;\;\;\;\;\; 0  & \mbox{elsewhere} \ ,
\end{array} \right. \label{S5Fz1}
 \ee


where $l_1=L/5-\Delta z_{12}$ and $l_2=L/5+\Delta z_{12}$ with
$\Delta z_{12}=L/50$; while $l_3=L/2-\Delta z_{34}$ and
$l_4=L/2+\Delta z_{34}$ with $\Delta z_{34}=L/10$; and, at last,
$l_5=4L/5-\Delta z_{56}$ and $l_6=4L/5+\Delta z_{56}$ with $\Delta
z_{56}=L/50$. In other words, the desired longitudinal shape, in
the range $0 \leq z \leq L$, is a parabolic function for $l_1 \leq
z \leq l_2$, a unitary step function for $l_3 \leq z \leq l_4$,
and again a parabola in the interval $l_5 \leq z \leq l_6$, being
zero elsewhere (within the interval $0 \leq z \leq L$, as we
said). In this example, let us put $L=0.2\;$m.

\h We can then easily calculate the coefficients $A_m$, which
appear in the superposition (\ref{S5soma3}), by inserting
eq.(\ref{S5Fz1}) into eq.(\ref{S5An}). Let us choose, for instance,
$Q=0.999\,\om_0/c$. This choice permits the maximum value $N_{\rm
max}=316$ for $m$, as one can infer from eq.(\ref{S5N}). Let us
emphasize that one is not compelled to use just $N=316$, but can
adopt for $N$ any values {\em smaller} than it; more generally,
any value smaller than that calculated via inequality (\ref{S5N}).
Of course, WHEN using the maximum value allowed for $N$, one gets a
better result.

\begin{figure}[!h]
\begin{center}
 \scalebox{1}{\includegraphics{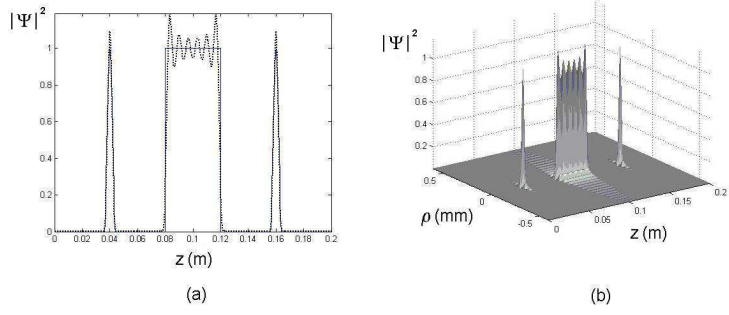}}
\end{center}
\caption{\textbf{(a)} Comparison between the intensity of the
desired longitudinal function $F(z)$ and that of our Frozen Wave
(FW), \ $\Psi(\rho=0,z,t)$, \ obtained from eq.(\ref{S5soma2}).
The solid line represents the function $F(z)$, and the dotted one
our FW. \textbf{(b)} 3D-plot of the field-intensity of the FW
chosen in this case by us.} \label{S5MHRFig1sec5}
\end{figure}

\h In the present case, let us adopt the value $N=30$. In
Fig.\ref{S5MHRFig1sec5}(a) we compare the intensity of the desired
longitudinal function $F(z)$ with that of the Frozen Wave, \
$\Psi(\rho=0,z,t)$, \ obtained from eq.(\ref{S5soma2}) by adopting
the mentioned value $N=30$.

\

\h One can verify that a good agreement between the desired
longitudinal behavior and our approximate FW is already got with
$N=30$. The use of higher values for $N$ can only improve the
approximation. \ Figure \ref{S5MHRFig1sec5}(b) shows the
3D-intensity of our FW, given by eq.(\ref{S5soma3}). One can
observe that this field possesses the desired longitudinal
pattern, while being endowed with a good transverse localization.

\

\emph{Second example} (controlling the transverse shape too):

\

We wish to take advantage of this example for addressing an
important question: \ We can expect that, for a desired
longitudinal pattern of the field intensity, by choosing smaller
values of the parameter $Q$ one will get FWs with narrower {\em
transverse} width [for the same number of terms in the series
entering eq.(\ref{S5soma3})], because of the fact that the Bessel
beams in eq.(\ref{S5soma3}) will possess larger transverse wave
numbers, and, consequently, higher transverse concentrations. \ We
can verify this expectation by considering, for instance, inside
the usual range $0 \leq z \leq L$, the longitudinal pattern
represented by the function

 \bb
 F(z) \ug \left\{\begin{array}{clr}
 -4\,\,\dis{\frac{(z-l_1)(z-l_2)}{(l_2 - l_1)^2}} \;\;\; &
 {\rm for}\;\;\; l_1 \leq z \leq l_2  \\

 \\
 \;\;\;\;\;\;\;\; 0  & \mbox{elsewhere}
\end{array} \right. \; , \label{S5Fz2}
 \ee

with $l_1=L/2-\Delta z$ and $l_2=L/2+\Delta z$.  Such a function
has a parabolic shape, with its peak centered at $L/2$ and with
longitudinal width $2 \Delta z/\sqrt{2}$.  By adopting $\lambda_0
= 0.632\;\mu$m (that is, $\om_0 = 2.98 \times 10^{15}\;$Hz), let
us use the superposition (\ref{S5soma3}) with {\em two} different
values of $Q$: \ We shall obtain two different FWs that, in spite
of having the same longitudinal intensity pattern, will possess
different transverse localizations. Namely, let us consider
$L=0.06\,$m and $\Delta z = L/100$, and the two values
$Q=0.999\,\om_0/c$ and $Q=0.995\,\om_0/c$. In both cases the
coefficients $A_m$ will be the same, calculated from
eq.(\ref{S5An}), using this time the value $N=45$ in superposition
(\ref{S5soma3}). The results are shown in Figs.3(a) and 3(b). Both
FWs have the same longitudinal intensity pattern, but the one with
the smaller $Q$ is endowed with a narrower transverse width.

\

\begin{figure}[!h]
\begin{center}
\scalebox{1}{\includegraphics{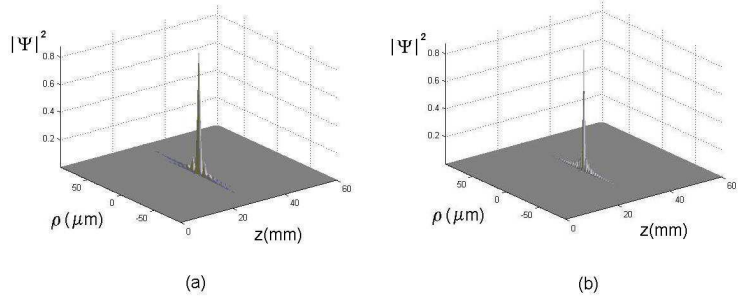}}
\end{center}
\caption{\textbf{(a)} The Frozen Wave with $Q=0.999\om_0/c$ and
$N=45$, approximately reproducing the chosen longitudinal pattern
represented by eq.(\ref{S5Fz2}). \ \textbf{(b)} A different Frozen
wave, now with $Q=0.995 \om_0/c$ (but still with $N=45$)
forwarding the same longitudinal pattern. We can observe that in
this case (with a lower value for $Q$) a higher transverse
localization is obtained.} \label{S5MHRFig3sec5}
\end{figure}

With this, we can get some control on the transverse spot size
through the parameter $Q$. Actually, eq.(\ref{S5soma3}), which
defines our FW, is a superposition of zero-order Bessel beams,
and, due to this fact, the resulting field is expected to possess
a transverse localization around $\rho=0$. Each Bessel beam in
superposition (\ref{S5soma3}) is associated with a central spot
with transverse size, or width, $\Delta\rho_m \approx
2.4/k_{\rho\,m}$. On the basis of the expected convergence of
series (\ref{S5soma3}), we can estimate the width of the
transverse spot of the resulting beam as being

\bb \Delta \rho \approx \frac{2.4}{k_{\rho\,m=0}} \ug
\frac{2.4}{\sqrt{\om_0^2/c^2 - Q^2}} \; , \label{S5tspot} \ee

which is the same value as that for the transverse spot of the
Bessel beam with $m=0$ in superposition (\ref{S5soma3}). Relation
(\ref{S5tspot}) can be useful: Once we have chosen the desired
longitudinal intensity pattern, \emph{we can choose even the size
of the transverse spot, and use relation (\ref{S5tspot}) for
evaluating the needed, corresponding value of parameter $Q$.}

For a more detailed analyzis concerning the spatial resolution and
residual intensity of the Frozen Waves, we refer the reader to
ref.[17].

\

\subsubsection{Increasing the control on the transverse shape by
using higher-order Bessel beams}

\

Here, we are going to argue that it is possible to increase even
more our control on the transverse shape by using higher-order
Bessel beams in our fundamental superposition (\ref{S5soma3}).

This new approach can be understood and accepted on the basis of
simple and intuitive arguments, which are not presented here, but
can be found in ref.[17]. A brief description of that approach follows
below.

The basic idea is obtaining the desired longitudinal intensity
pattern not along the axis $\rho=0$, but on a cylindrical surface
corresponding to $\rho=\rho'>0$.

To do this, we first proceed as before: Once we have chosen the
desired longitudinal intensity pattern $F(z)$, within the interval
$0 \leq z \leq L$, we calculate the coefficients $A_m$ as before,
i.e., \ $A_m = (1/L) \int_{0}^{L}\,F(z)\,{\rm exp}(-i 2 \pi m z
/L)\,\drm z$, and  $k_{\rho\,m} = \sqrt{\om_0^2 - \left(Q + 2 \pi
m/L \right)^2}$.

Afterwards, we just replace the zero-order Bessel beams
$J_0(k_{\rho\,m}\rho)$, in superposition (\ref{S5soma3}), with
higher-order Bessel beams, $J_{\mu}(k_{\rho\,m}\rho)$, to get

\bb \dis{\Psi(\rho,z,t) \ug
e^{-i\,\om_0\,t}\,e^{i\,Q\,z}\,\sum_{m=-N}^{N}
A_m\,J_{\mu}(k_{\rho\,m}\,\rho)\,e^{i\,\frac{2\pi}{L}m\,z} } \; ,
\label{S5soma4} \ee

With this, and based on intuitive arguments[17], we can expect
that the desired longitudinal intensity pattern, initially
constructed for $\rho=0$, will approximately shift to $\rho =
\rho'$, where $\rho'$ represents the position of the first maximum
of the Bessel function, i.e the first positive root of the
equation \
$(\drm\,J_{\mu}(k_{\rho\,m=0}\,\rho)/\drm\rho)|_{\rho'}=0$.

By such a procedure, one can obtain very interesting stationary
configurations of field intensity, as ``donuts", cylindrical
surfaces, and much more.

In the following example, we show how to obtain, e.g., a
cylindrical surface of stationary {\em light}. \ To get it, within
the interval $0 \leq z \leq L$, let us first select the
longitudinal intensity pattern given by eq.(\ref{S5Fz2}), with
$l_1=L/2-\Delta z$ and $l_2=L/2+\Delta z$, and with $\Delta z =
L/300$. Moreover, let us choose $L=0.05\,$m, $Q=0.998\,\om_0/c$,
and use $N=150$.

\h Then, after calculating the coefficients $A_m$ by
eq.(\ref{S5An}), we have recourse to superposition
(\ref{S5soma4}). In this case, we choose $\mu =4$. \ According to
the previous discussion, one can expect the desired longitudinal
intensity pattern to appear shifted to $\rho ' \approx
5.318/k_{\rho\,m=0}=8.47\,\mu$m, where 5.318 is the value of
$k_{\rho\,m=0}\,\rho$ for which the Bessel function
$J_{4}(k_{\rho\,m=0}\,\rho)$ assumes its maximum value, with
$k_{\rho\,m=0} = \sqrt{\om_0^2 - Q^2}$. \ The figure
\ref{S5MHRFig3sec5} below shows the resulting intensity field.

In Fig.\ref{S5MHRFig3sec5}(a) the transverse section of the
resulting beam for $z=L/2$ is shown. The transverse peak intensity
is located at $\rho=7.75\,\mu$m, with a $8.5\%$ difference w.r.t.
the predicted value of $8.47\,\mu$m. Figure \ref{S5MHRFig3sec5}(b)
shows the orthogonal projection of the resulting field, which
corresponds to nothing but a cylindrical surface of stationary
light (or other fields).

\

\begin{figure}[!h]
\begin{center}
\scalebox{.6}{\includegraphics{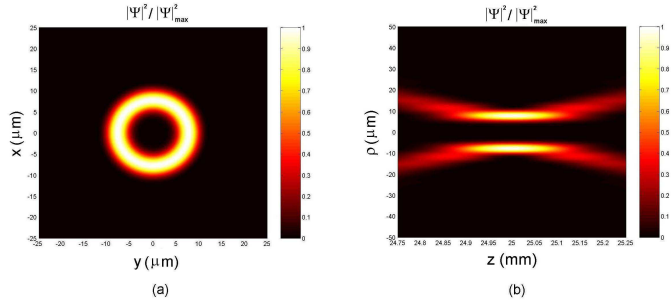}}
\end{center}
\caption{\textbf{(a)} Transverse section at $z=L/2$ of the
considered higher-order FW. \ \textbf{(b)} Orthogonal projection
of the three-dimensional intensity pattern of the same
higher-order FW.} \label{S5MHRFig3sec5}
\end{figure}

We can see that the desired longitudinal intensity pattern has
been approximately obtained, but, as wished, shifted from $\rho=0$
to $\rho = 7.75\,\mu$m, and the resulting field resembles a
cylindrical surface of stationary light with radius $7.75\,\mu$m
and length $238\,\mu$m. Donut-like configurations of light (or
sound) are also possible.

\

\subsection{Stationary wavefields
with arbitrary longitudinal shape in absorbing media: Extending
the method.}

\

When propagating in a non-absorbing medium, the so-called
nondiffracting waves maintain their spatial shape for long
distances. However, the situation is not the same when dealing
with absorbing media. In such cases, both the ordinary and the
nondiffracting beams (and pulses) will suffer the same effect: an
exponential attenuation along the propagation axis.

Here, we are going to make an extension[10] of the method given
above to show that, through suitable superpositions of
equal-frequency Bessel beams, it is possible to obtain
nondiffracting beams in {\it absorbing media}, whose longitudinal
intensity pattern can assume any desired shape within a chosen
interval $ 0 \leq z \leq L$ of the propagation axis $z$.

As a particular example, we obtain new nondiffracting beams
capable to resist the loss effects, maintaining amplitude and spot
size of their central core for long distances.

It is important to stress that in this new method there is no
active participation of the material medium. Actually, the energy
absorption by the medium continues to occur normally, the
difference being in the fact that these new beams have an initial
transverse field distribution, such to be able to reconstruct
(even in the presence of absorption) their central cores for
distances considerably longer than the penetration depths of
ordinary (nondiffracting or diffracting) beams. In this sense, the
present method can be regarded as extending, for absorbing media,
the self-reconstruction properties[27] that usual Localized Waves
are known to possess in loss-less media.

In the same way as for lossless media, we construct a Bessel beam
with angular frequency $\om$ and axicon angle $\theta$ in the
absorbing materials by superposing plane waves, with the same
angular frequency $\om$, and whose wave vectors lie on the surface
of a cone with vertex angle $\theta$. The refractive index of the
medium can be written as $n(\om) = n_R(\om) + in_I(\om)$, quantity
$n_R$ being the real part of the complex refraction index and
$n_I$ the imaginary one, responsible for the absorbtion effects.
With a plane wave, the penetration depth $\delta$ for the
frequency $\om$ is given by $\delta = 1/\alpha= c/2\om n_I$, where
$\alpha$ is the absorption coefficient.

In this way, a zero-order Bessel beam in dissipative media can be
written as \ $\psi = J_0(k_{\rho}\rho){\rm exp}(i\beta z){\rm
exp}(-i\om t)$ with $\beta = n(\om)\om \cos\theta/c = n_R\om
\cos\theta/c + in_I\om \cos\theta/c \equiv \beta_R + i\beta_I$;
$k_{\rho}=n_R\om \sin\theta/c + in_I\om \sin\theta/c \equiv
k_{\rho R} + ik_{\rho I}$, and so $k_{\rho}^2 = n^2\om^2/c^2 -
\beta^2$. \ In this way,] it results \  $\psi=J_0((k_{\rho R} + ik_{\rho
I})\rho){\rm exp}(i\beta_R z){\rm exp}(-i\om t){\rm exp}(-\beta_I
z)$, \  where $\beta_R$, $k_{\rho R}$ are the real parts of the
longitudinal and transverse wave numbers, and $\beta_I$, $k_{\rho
I}$ are the imaginary ones, while the absorption coefficient of a
Bessel beam with axicon angle $\theta$ is given by
$\alpha_{\theta}=2\beta_I=2n_I\om \cos\theta/c$, its penetration
depth being $\delta_{\theta}=1/\alpha_{\theta}=c/2\om
n_I\cos\theta$.

Due to the fact that $k_{\rho}$ is complex, the amplitude of the
Bessel function $J_0(k_{\rho}\rho)$ starts decreasing from
$\rho=0$ till the transverse distance $\rho=1/2k_{\rho I}$, and
afterwards it starts growing exponentially.  This behavior is not
physically acceptable, but one must remember that it occurs only
because of the fact that an ideal Bessel beam needs an infinite
aperture to be generated. However, in any real situation, when a
Bessel beam is generated by finite apertures, that exponential
growth in the transverse direction, starting after
$\rho=1/2k_{\rho I}$, will {\it not} occur indefinitely, stopping
at a given value of $\rho$. Let us moreover emphasize that, when
generated by a finite aperture of radius $R$, the truncated Bessel
beam[17] possesses a depth of field $Z = R/\tan\theta$, and can
be approximately described by the solution given in the previous
paragraph, for $\rho<R$ and $z < Z$.

Experimentally, to guarantee that the mentioned exponential growth
in the transverse direction {\it does not} even start, so as to
meet only a decreasing transverse intensity, the radius $R$ of the
aperture used for generating the Bessel beam should be $R \leq
1/2k_{\rho I}$. However, as noted by Durnin et al., the same
aperture has to satisfy also the relation $R \geq 2\pi/k_{\rho
R}$. From these two conditions, we can infer that, in an absorbing
medium, a Bessel beam with just a decreasing transverse intensity
can be generated only when the absorption coefficient is $ \alpha
< 2/\lambda$, i.e., if the penetration depth is $\delta > \lambda
/2$. The method developed in this paper does refer to these cases,
i.e. we can always choose a suitable finite aperture size in such
a way that the truncated versions of all solutions presented in
this work, including the general one given by eq.(6), will not
present any unphysical behavior. Let us now outline our method.

Consider an absorbing medium with the complex refraction index
$n(\om) = n_R(\om) + in_I(\om)$, and the following superposition
of $2N + 1$ Bessel beams with the same frequency $\om$:

\bb
\begin{array}{clr}
\Psi(\rho,z,t) = \dis{\sum_{m=-N}^{N}} A_m\,J_0\left((k_{\rho R_m}
+ ik_{\rho
I_m})\rho\right)\,e^{i\,\be_{R_m}z}\,e^{-i\,\om\,t}\,e^{-\be_{I_m}z}
\; ,
\end{array} \label{S52soma1} \ee
\ where the $m$ are integer numbers, the $A_m$ are constant (yet unknown)
coefficients, quantities $\be_{R_m}$ and $k_{\rho R_m}$ ($\be_{I_m}$ and
$k_{\rho I_m}$) are the real (the imaginary) parts of the
complex longitudinal and transverse wave numbers of the $m$-th
Bessel beam in superposition (\ref{S52soma1}); the following
relations being satisfied

\bb
 k_{\rho_m}^2 \ug n^2\frac{\om^2}{c^2} - \be_{m}^2 \label{S52kr}
\ee

\bb \frac{\be_{R_m}}{\be_{I_m}} \ug \frac{n_R}{n_I} \label{S52bei}
\ee \ where $\be_m = \be_{R_m} + i\be_{I_m}$, $k_{\rho_m} =
k_{\rho R_m} + ik_{\rho I_m}$, with $k_{\rho R_m}/k_{\rho
I_m}=n_R/n_I$.

Our goal is now to find out the values of the longitudinal wave
numbers $\be_m$ and the coefficients $A_m$ in order to reproduce
approximately, inside the interval $0 \leq z \leq L$ (on the axis
$\rho=0$), a {\it freely chosen} longitudinal intensity pattern
that we call $|F(z)|^2$.

The problem for the particular case of lossless media[16,17],
i.e., when $n_I=0 \rightarrow \be_{I_m}=0$, was solved in the
previous subsection. For those cases, it was shown that the choice
$\beta=Q+2\pi m/L$, with $A_m = \int_{0}^{L} F(z){\rm exp}(-i2\pi
mz/L)/L\,\,dz$ can be used to provide approximately the desired
longitudinal intensity pattern $|F(z)|^2$ on the propagation axis,
within the interval $0\leq z \leq L$, and, at the same time, to
regulate the spot size of the resulting beam by means of the
parameter $Q$, which parameter can be also used to obtain large field depths
and also to inforce the linear polarization approximation to the
electric field for the TE electromagnetic wave (see details in
refs.[16,17]).

However, when dealing with absorbing media, the procedure
described in the last paragraph does not work, due to the presence
of the functions $\exprm (-\be_{I_m}z)$ in the superposition
(\ref{S52soma1}), since in this case that series does not became a
Fourier series when $\rho=0$.

On attempting to overcome this limitation, let us write the real
part of the longitudinal wave number, in superposition
(\ref{S52soma1}), as

\bb \be_{R_m} \ug Q + \frac{2\pi m}{L} \label{S52br} \ee

with

\bb 0 \leq Q + \frac{2\pi m}{L} \leq n_R \frac{\om}{c}
\label{S52cond} \ee \ where this inequality guarantees forward
propagation only, with no evanescent waves.

In this way the superposition (\ref{S52soma1}) can be written

\bb
\begin{array}{clr}
\Psi(\rho,z,t) = e^{-i\,\om\,t}\,e^{i\,Qz}\, \dis{\sum_{m=-N}^{N}}
A_m\,J_0\left((k_{\rho R_m} + ik_{\rho I_m})\rho\right)
\,e^{i\,\frac{2\pi m}{L}z}\,e^{-\be_{I_m}z}  \; ,
\end{array} \label{S52soma2} \ee
\ where, by using (\ref{S52bei}), we have $\be_{I_m}=(Q + 2\pi
m/L)n_I/n_R$, and $k_{\rho_m}=k_{\rho R_m} + ik_{\rho I_m}$ is
given by (\ref{S52kr}). Obviously, the discrete superposition
(\ref{S52soma2}) could be written as a continuous one (i.e., as an
integral over $\be_{R_m}$) by taking $L \rightarrow \infty$, but
we prefer the discrete sum due to the difficulty of obtaining
closed-form solutions to the integral form.

Now, let us examine the imaginary part of the longitudinal wave
numbers. The minimum and maximum values among the $\be_{I_m}$ are
$(\be_I)_{\rm min}=(Q-2\pi N/L)n_I/n_R$ and $(\be_I)_{\rm
max}=(Q+2\pi N/L)n_I/n_R$, the central one being given by
$\overline{\be}_I \equiv (\be_I)_{m=0} = Q n_I/n_R $. With this in
mind, let us evaluate the ratio $\Delta = [(\be_I)_{\rm max} -
(\be_I)_{\rm min}]/{\overline{\be}_I} = 4 \pi N/LQ$.

Thus, when $\Delta <<1$, there are no considerable differences
among the various $\be_{I_m}$, since $\be_{I_m} \approx
\overline{\be}_I$ holds for all $m$. In the same way, there are no
considerable differences among the exponential attenuation
factors, since $\exprm (-\be_{I_m}z) \approx \exprm
(-\overline{\be}_I z)$. So, when $\rho=0$ the series in the r.h.s.
of eq.(\ref{S52soma2}) can be approximately considered a truncated
Fourier series {\it multiplied by} the function $\exprm
(-\overline{\be}_I z)$ and, therefore, superposition
(\ref{S52soma2}) can be used to reproduce approximately the
desired longitudinal intensity pattern $|F(z)|^2$ (on $\rho=0$),
within $0\leq z \leq L$, when the coefficients $A_m$ are given by

\bb A_m \ug \frac{1}{L}\,\int_{0}^{L} F(z)\,e^{\overline{\be}_I
z}e^{-i\,\frac{2\pi m}{L}z}\,dz ,\label{S52am} \ee being necessary
the presence of the of the factor $\exprm (\overline{\be}_I z)$ in
the integrand to compensate for the factors $\exprm (-\be_{I_m}z)$
in superposition (\ref{S52soma2}).

Since we are adding together zero-order Bessel functions, we can
expect a good field concentration around $\rho=0$.

In short, we have shown in this Section how one can get, in an
{\it absorbing medium}, a {\it stationary} wave-field with a good
transverse concentration, and whose longitudinal intensity pattern
(on $\rho=0$) can approximately assume {\it any desired shape}
$|F(z)|^2$ within the predetermined interval $0 \leq z \leq L$.
The method is a generalization of a previous one[16,17] and
consists in the superposition of Bessel beams in EQ.(\ref{S52soma2}),
the real and imaginary parts of their longitudinal wave numbers
being given by eqs.(\ref{S52br})and (\ref{S52bei}), while their
complex transverse wave numbers are given by eq.(\ref{S52kr}),
and, finally, the coefficients of the superposition are given by
eq.(\ref{S52am}). The method is justified, since $4\pi N/LQ << 1$;
happily enough, this condition is satisfied in a great number of
situations.

Regarding the generation of these new beams, given an apparatus
capable of generating a single Bessel beam, we can use an array of
such apparatuses to generate a sum of them, with the appropriate
longitudinal wave numbers and amplitudes/phases [as required by
the method], thus producing the desired beam. For instance, we can
use[16,17] a laser illuminating an array of concentric annular
apertures (located at the focus of a convergent lens) with the
appropriate radii and transfer functions, able to yield both the
correct longitudinal wave numbers (once a value for $Q$ has been
chosen) and the coefficients $A_n$ of the fundamental
superposition (\ref{S52soma2}).

\subsubsection{Some Examples}

For generality's sake, let us consider a hypothetical medium in
which a typical XeCl excimer laser ($\lambda = 308 {\rm nm}
\rightarrow \om = 6.12\times 10^{15}$Hz) has a penetration depth
of 5 cm; i.e. an absorption coefficient $\alpha = 20 {\rm
m}^{-1}$, and therefore $n_I = 0.49\times 10^{-6} $. Besides this,
let us suppose that the real part of the refraction index for this
wavelength is $n_R = 1.5$ and therefore $n = n_R + in_I = 1.5 +
i\,0.49\times 10^{-6}$. Note that the value of the real part of
the refractive index is not so important for us, since we are
dealing with monochromatic wave fields.

A Bessel beam with $\om  = 6.12\times 10^{15}$Hz and with an
axicon angle $\theta = 0.0141 \;$rad (so, with a transverse spot of
radius $8.4\,\mu$m), when generated by an aperture, say, of radius
$R=3.5\;$mm, can propagate in vacuum a distance
equal to $Z=R/\tan\theta=25\;$cm while resisting the diffraction
effects. However, in the material medium considered here, the
penetration depth of this Bessel beam would be only $z_p = 5\;$cm.
Now, let us set forth two interesting applications of the method.

\

\textbf{First Example}: Almost Undistorted Beams in Absorbing
Media.

\

Now, we can use the extended method to obtain, in the same medium
and for the same wavelength, an almost undistorted beam capable of
preserving its spot size and the intensity of its {\it central
core} for a distance many times larger than the typical
penetration depth of an ordinary beam (nondiffracting or not).

With this purpose, let us suppose that, for this material medium,
we wish a beam (with $\om  = 6.12\times 10^{15}$Hz) that maintains
amplitude and spot size of its central core for a distance of
$25\;$cm, i.e. a distance 5 times greater than the penetration
depth of an ordinary beam with the same frequency. We can model
this beam by choosing the desired longitudinal intensity pattern
$|F(z)|^2$ (on $\rho=0$), within $0 \leq z \leq L$,

\bb
 F(z) \ug \left\{\begin{array}{clr}
&1 \;\;\; {\rm for}\;\;\; 0 \leq z \leq Z  \\

&0 \;\;\; \mbox{elsewhere} ,
\end{array} \right.  \label{S52Fz}
 \ee
and by putting $Z=25\;$cm, with, for example, $L=33\;$cm.

Now, the Bessel beam superposition (\ref{S52soma2}) can be used to
reproduce approximately this intensity pattern, and to this
purpose let us choose $Q=0.9999\om / c$ for the $\be_{R_m}$ in
(\ref{S52br}), and $N=20$ (note that, according to inequality
(\ref{S52cond}), $N$ could assume a maximum value of $158$.)

Once we have chosen the values of $Q$, $L$ and $N$, the values of
the complex longitudinal and transverse Bessel beams wave numbers
happen to be defined by relations (\ref{S52br}), (\ref{S52bei})
and (\ref{S52kr}). Eventually, we can use eq.(\ref{S52am}) and
find out the coefficients $A_m$ of the fundamental superposition
(\ref{S52soma2}), that defines the resulting stationary
wave-field.

Let us just note that the condition $4\pi N/LQ << 1$ is perfectly
satisfied in this case.

\begin{figure}[!h]
\begin{center}
\scalebox{.8}{\includegraphics{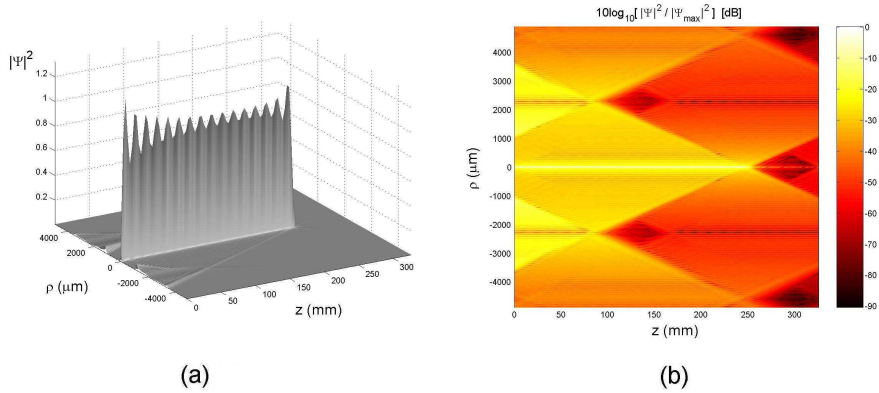}}
\end{center}
\caption{\textbf{(a)} Three-dimensional field-intensity of the
resulting beam. \textbf{(b)} The resulting beam, in an orthogonal
projection and in {\bf logaritmic} scale.} \label{S52MHRFig5sec5}
\end{figure}

In Fig. (\ref{S52MHRFig5sec5})(a) we can see the 3D
field-intensity of the resulting beam. One can see that the field
possesses a good transverse localization (with a spot size smaller
than $10 \; \mu$m), it being capable of maintaining spot size and
{\it intensity} of its central core till the desired distance (a
better result could be reached by using a higher value of $N$).

\

It is interesting to note that at this distance (25 cm), an
ordinary beam would have got its initial field-intensity
attenuated $148$ times.

As we have said in the Introduction, the energy absorption by the
medium continues to occur normally; the difference is that these
new beams have an initial transverse field distribution
sophisticated enough to be able to reconstruct (even in the
presence of absorption) their central cores till a certain
distance. For a better visualization of this field-intensity
distribution and of the energy flux, Fig.(\ref{S52MHRFig5sec5})(b)
shows the resulting beam, in an
orthogonal projection and in {\bf logarithmic} scale. It is clear
that the energy comes from the lateral regions, in order to
reconstruct the central core of the beam. On the plane $z=0$,
within the region $\rho \leq R=3.5\,$mm, there is a uncommon field
intensity distribution, being very dispersed instead of
concentrated. This uncommon initial field intensity distribution
is responsible for constructing the central core of the resulting
beam and for its reconstruction till the distance $z=25\,$cm. Due
to absorption, the beam (total) energy flowing through
different $z$ planes, is not constant, but the energy flowing
IN the beam spot area and the beam spot size itself are
conserved till (in this case) the distance $z=25\,$ cm.

\

\textbf{Second Example}: Beams in absorbing media with a growing
longitudinal field intensity.

\

Considering again the previous hypothetical medium, in which an
ordinary Bessel beam with $\theta = 0.0141 \; $rad and $\om  =
6.12\times 10^{15}$Hz has a penetration depth of $5\;$cm, we aim
to construct now a beam that, instead of possessing a {\it
constant} core-intensity till the position $z=25\;$cm, presents
on the contrary a (moderate) exponential {\em growth} of its
intensity, till that distance ($z=25\;$cm).

Let us assume we wish to get the longitudinal intensity pattern
$|F(z)|^2$, in the interval $0<z<L$,

\bb
 F(z) \ug \left\{\begin{array}{clr}
& {\rm exp}(z/Z) \;\;\; {\rm for}\;\;\; 0 \leq z \leq Z  \\

&0 \;\;\; \mbox{elsewhere} ,
\end{array} \right.  \label{S52Fz2}
 \ee
with $Z=25\,$cm and $L=33\;$cm.

Using again $Q=0.9999\om / c$, $N=20$, we CAN proceed as in the first
example, calculating the complex longitudinal and transverse
Bessel beams wave numbers and finally the coefficients $A_m$ of
the fundamental superposition (\ref{S52soma2}).

\begin{figure}[!h]
\begin{center}
\scalebox{.8}{\includegraphics{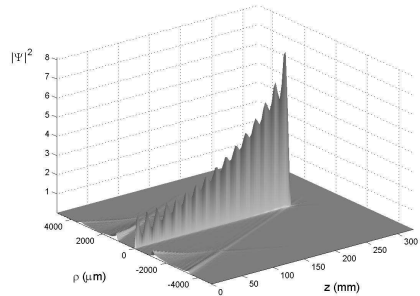}}
\end{center}
\caption{Three-dimensional field-intensity of the resulting beam,
in an absorbing medium, with a growing longitudinal field
intensity.} \label{S52MHRFig6sec5}
\end{figure}

In Fig. (\ref{S52MHRFig6sec5}) we can see the 3D field-intensity
of the resulting beam. One can see that the field presents the
desired longitudinal intensity pattern with a good transverse
localization (a spot size smaller than $10 \; \mu$m).

\

Obviously, the amount of energy necessary to construct these new
beams is greater than that necessary to generate an ordinary beam
in a non-absorbing medium. And it is also clear that there is a
{\it limitation} on the depth of field of these new beams. In the
first example, for distances longer than 10 times the penetration
depth of an ordinary beam, besides a greater energy demand, we
meet the fact that the field-intensity in the lateral regions
would be even higher than that of the core, and the field would
loose the usual characteristics of a beam (transverse field
concentration)

\


\

\

References

\

[1] A much more complete list of references can be found in
the previous chapter (Chapter 1) of this book, that is, of {\em Localized
Waves}, ed. by H.E.Hern\'andez-Figueroa, M.Zamboni-Rached and E.Recami
(J.Wiley; in press).

[2] J. N. Brittingham, ``Focus wave modes in homogeneous Maxwell's
equations: transverse    electric mode,'' J. Appl. Phys., Vol. 54,
pp. 1179-1189 (1983).

[3] J. Durnin, J. J. Miceli e J. H. Eberly, ``Diffraction-free
beams,'' Phys. Rev. Lett., Vol. 58, pp. 1499-1501 (1987).

[4] J.-y. Lu, J. F. Greenleaf, ``Nondiffracting X-waves - Exact
solutions to free-space wave equation and their finite aperture
realizations," IEEE Transactions in Ultrasonics Ferroelectricity
and Frequency Control, Vol.39, pp.19-31 (1992); and refs. therein.

[5] H.S\~{o}najalg, P.Saari, ``Suppression of temporal spread of
ultrashort pulses in dispersive media by Bessel beam generators'',
Opt. Letters, Vol. 21, pp.1162-1164 (1996).

[6] M. Zamboni-Rached, K. Z. N\'obrega, H. E. Hern\a'ndez-Figueroa e
E. Recami, ``Localized Superluminal solutions to the wave equation
in (vacuum or) dispersive media, for arbitrary frequencies and
with adjustable bandwidth'', Optics Communications, Vol.226, pp.
15-23 (2003).

[7] M.A.Porras, R.Borghi, and M.Santarsiero, ``Suppression of
dispersion broadening of light pulses with Bessel-Gauss beams",
Optics Communications, Vol. 206, 235-241 (2003).

[8] C.Conti, S.Trillo, P.Di Trapani, G.Valiulis, A.Piskarskas,
O.Jedrkiewicz and J.Trull, ``Nonlinear electromagnetic X-waves",
Phys. Rev. Lett., vol.90, paper no.170406, May 2003; and refs.
therein.

[9] J. Salo, J. Fagerholm,  A. T. Friberg,  and M. M. Salomaa,
``Nondiffracting Bulk-Acoustic X waves in Crystals,'' Phys. Rev.
Lett., Vol. 83, pp. 1171-1174 (1999); and refs. therein.

[10]  M. Zamboni-Rached, "Diffraction-Attenuation resistant beams
in absorbing media", Optics Express, Vol. 14, pp.1804-1809 (2006).

[11] I. M. Besieris, A. M. Shaarawi e R. W. Ziolkowski, ``A
bidirectional traveling plane wave representation of exact
solutions of the scalar wave equation'', J. Math. Phys., Vol. 30,
pp. 1254-1269 (1989).

[12] M. Zamboni-Rached, `` Localized Waves: Structure and
Applications," M.Sc. Thesis (Physics Department, Unicamp, 1999).

[13] M. Zamboni-Rached,  E. Recami, H. E. Hern\'andez-Figueroa, "New
localized Superluminal solutions to the wave equations-with finite
total energies and arbitrary frequencies", European Physics
Journal D, Vol. 21, 217-228 (2002).

[14] M.Zamboni-Rached, ``Localized waves in diffractive/dispersive
media", PhD Thesis, Aug.2004, Universidade Estadual de Campinas,
DMO/FEEC. Download at:

http://libdigi.unicamp.br/document/?code=vtls000337794

[15] S. Longhi, ``Localized subluminal envelope pulses in
dispersive media," Optics Letters, Vol. 29, pp.147-149 (2004);and
refs. therein.

[16] M. Zamboni-Rached, ``Stationary optical wave fields with
arbitrary longitudinal shape by superposing equal frequency Bessel
beams: Frozen Waves'', Optics Express, Vol. 12, pp.4001-4006
(2004).

[17] M. Zamboni-Rached, E. Recami, H. E. Hern\'andez-Figueroa,
``Theory of Frozen Waves: Modelling the Shape of Stationary Wave
Fields", Journal of Optical  Society of America A, Vol. 22,
pp.2465-2475.

[18] S. V. Kukhlevsky and M. Mechler, ``Diffraction-free
subwavelength-beam optics at nanometer scale," Optics
Communications, Vol. 231, pp.35-43 (2004).

[19] I.S.Gradshteyn, and I.M.Ryzhik: Integrals, Series and
Products, 4th edition (Acad. Press; New York, 1965).

[20] A.M.Shaarawi, I.M.Besieris and T.M.Said, ``Temporal focusing
by use of composite X-waves", J. Opt. Soc. Am., A, vol.20,
pp.1658-1665, Aug.2003.

[21] M. Zamboni-Rached, A. Shaarawi, E. Recami, ``Focused X-Shaped
Pulses", Journal of Optical  Society of America A, Vol. 21, pp.
1564-1574 (2004).

[22] M. Zamboni-Rached, H.E. Hern\'andez-Figueroa, E. Recami
``Chirped Optical X-type Pulses," Journal of Optical  Society of
America A, Vol. 21, pp. 2455-2463 (2004).

[23] G. Agrawal: Nonlinear Fiber Optics (Academic Press; 4th
edition, 2006).

[24] Z. Bouchal and M. Olivik, ```Non-diffractive vector Bessel
beams," Journal of Modern Optics, Vol. 42, pp.1555-1566 (1995).

[25] E.Recami: ``On localized X-shaped Superluminal solutions to
Maxwell equations," Physica A, Vol. 252, pp.586-610 (1998).

[26] Z. Bouchal, ``Controlled spatial shaping of nondiffracting
patterns and arrays", Optics Letters, Vol. 27, pp. 1376-1378
(2002).

[27] R. Grunwald, U. Griebner, U. Neumann, and V. Kebbel,
Self-reconstruction of ultrashort-pulse Bessel-like X-waves, CLEO
/ QELS 2004, San Francisco, May 16-21, 2004, Conference Digest
(CD-ROM), paper number CMQ7.

\end{document}